\documentclass[preprint]{aastex}

\shorttitle{Stellar Cluster Candidates}
\shortauthors{Richards et al.}

\begin{document}

\title{Multiwavelength Observations of Massive Stellar Cluster Candidates in the Galaxy\footnotemark\footnotetext{Accepted in the Astronomical Journal; to be published Fall 2012}}
\author{Emily E. Richards\footnotemark\footnotetext{Current affiliation: Department of Astronomy, Swain West, Indiana University, Bloomington, IN 47405}, Cornelia C. Lang}
\affil{Department of Physics \& Astronomy, Van Allen Hall, University of Iowa, Iowa City, IA 52242}
\email{er7@indiana.edu}

\author{Christine Trombley, Donald F. Figer} 
\affil{Center for Detectors, Rochester Institute of Technology, Rochester, NY 14623} 

\begin{abstract}

The Galaxy appears to be richer in young, massive stellar clusters than previously known, due to advances in infrared surveys which have uncovered deeply embedded regions of star formation. Young, massive clusters can significantly impact the surrounding interstellar medium (ISM) and hence radio observations can also be an important tracer of their activity. Several hundred cluster candidates are now known by examining survey data. Here we report on multiwavelength observations of six of these candidates in the Galaxy. We carried out 4.9 and 8.5 GHz VLA observations of the radio emission associated with these clusters to obtain the physical characteristics of the surrounding gas, including the Lyman continuum photon flux and ionized gas mass. {\it Spitzer} Infrared Array Camera observations were also made of these regions, and provide details on the stellar population as well as the dust continuum and polycyclic aromatic hydrocarbon emission. When compared to the known young, massive clusters in the Galaxy, the six cluster candidates have less powerful Lyman ionizing fluxes and ionize less of the H II mass in the surrounding ISM. Therefore, these cluster candidates appear to be more consistent with intermediate-mass clusters (10$^3$-10$^4$ M$_\sun$).

\end{abstract}


\clearpage

\section{INTRODUCTION}

Massive stars return a significant fraction of their mass to the interstellar medium (ISM) by means of stellar winds and supernovae. Massive stars are rare, have short lifetimes, and are predominantly observed in young, massive ($\le$ 20 Myr, $\ge$ 10$^4$ M$_\sun$) clusters. To date, few such clusters are known and well-studied in the Galaxy because many are located in the highly obscured Galactic disk. A complete census of young, massive stellar clusters in the Galaxy is needed to determine their total number and distribution. Estimates have shown that the Galaxy should host close to 100 of these clusters (\citealt{H-P08}). However, only a few such clusters are well known and studied. For example, the Arches, Quintuplet, and Central clusters in the Galactic center were uncovered with infrared spectroscopic observations, and are responsible for heating and shaping much of the surrounding ISM (\citealt{Cotera96}; \citealt{Figer99a}b; \citealt{Lang97}, \citeyear{Lang01a}, \citeyear{Lang02}). In addition, several other young, massive clusters in the Galaxy have been recently studied: RSGC 1 \&~2 (\citealt{Davies08}), W49a (\citealt{H-A05,de Pree97}), NGC 3603 (\citealt{Melena08,Borissova08,de Pree99}), and Westerlund 1 (\citealt{Brandner08,K-D07}).

\paragraph{}Over the last decade, surveys in the near-infrared and mid-infrared (2MASS and {\it Spitzer}/GLIMPSE in particular) have revealed as many as 1000 candidate clusters (e.g. \citealt{Bica03,Mercer05}). A sample of 40 possible young, massive clusters were selected from these candidates using multiwavelength data, including radio and optical observations. Among these 40 sources, six were identified as having bright associated radio emission from the NRAO VLA Sky Survey (VLA-NVSS; \citealt{Condon98}). Here, we present a multiwavelength study of these six cluster candidates in the Galaxy. The radio observations in this study have higher sensitivity and resolution than the VLA-NVSS data, and provide greater insight into the properties of the gas and dust surrounding these clusters. The primary goal of the radio data is to determine the properties of the ISM surrounding the candidate massive clusters, and to compare them to the known young, massive clusters in the Galaxy. Estimates of the ionized mass and ionizing flux can be obtained with the radio data, from which the candidate sources can be categorized as supermassive or intermediate-mass clusters. The primary goal of the infrared data is to characterize the infrared emission surrounding the clusters as well as the infrared emission arising from cluster members. Comparison between the radio and infrared observations provides additional insight about the interstellar environment of the stellar clusters and the impact of the stars on the surrounding medium.

\section{CANDIDATE CLUSTERS: BACKGROUND INFORMATION}

Presented here is a summary of the background information for the six candidate massive stellar clusters. Table \ref{tbl-1} lists the six candidate sources, including the most recent distance estimates and alternative identifiers. Note that the names used here have been shortened from names that originate in the Bica \&~Dutra catalogs of infrared star clusters (\citealt{Bica03,Dutra03,D-B01}). Sharpless (Sh2) numbers are given for the clusters that have been observed in the Sharpless catalog of H II regions (\citealt{Sharpless59}). 


\subsection{BD52}

BD52 is a stellar cluster located in the H II region Sh2-187. \cite{Russeil07} used the spectrophotometric distance of a single star (B2.5V) to determine a distance of 1.44 ${\pm}$ 0.26 kpc to the stellar cluster. According to a multifrequency study by \cite{Joncas92}, the H II region is fairly young with a kinematical age of $\sim$2$\times$10$^{5}$ yr. Optical and 1.4 GHz radio emission observed in this study suggest BD52 is enshrouded in its parental cloud, and that a B0 zero-age-main-sequence star is the exciting source. The presence of a water maser demonstrated by \cite{Zinchenko98} provides evidence that this H II region could be undergoing star formation.

\subsection{BD65}

BD65 is a stellar cluster located in the H II region Sh2-209. \cite{Klein05} have studied this complex at millimeter wavelengths to look for evidence of embedded star formation. Their observations show that the region around BD65 can be described as a ring of cloud cores around a star cluster, and the complex may be a site of triggered star formation. \cite{Klein05} adopted a distance of 8.2 kpc, yet there is some discrepancy in the distance to this source (estimates range from 4.9 to 9.8 kpc; \citealt{Bica03,Caplan00}). A distance to BD65 of 9.8 kpc is adopted in this work based on the kinematic and photometric distance provided by \cite{Caplan00}.

\subsection{BD84}

This appears to be one of the first times BD84 has been thoroughly studied. However, it has been observed in several H II region surveys and has been associated with the H II region Sh2-283. Its distance has been determined spectrophotometrically to be 8.06 ${\pm}$ 0.30 kpc by \cite{Russeil07}. 

\subsection{BD95}

The Galactic H II region Sh2-294, or BD95, has been studied in several papers, mostly H II region surveys, including a multiwavelength study of the region (\citealt{Samal07}). The distance to BD95 has been estimated several times; most recently it has been determined spectrophotometrically to be 3.24 ${\pm}$ 0.56 kpc (\citealt{Russeil07}). The ionizing star is believed to be a star of spectral type $\sim$B0V (\citealt{Samal07}). \cite{Yun08} confirm the ionizing source, and quote a spectral type of B0.5V. Their high resolution near infrared images reveal a double cluster morphology with one cluster fully embedded and optically invisible. 

\subsection{DB7}

DB7 is associated with an H II region known as Sh2-307. To the SW of DB7 is an ultracompact H II region and photodissociation region (PDR) associated with the DB8 stellar cluster (\citealt{Messineo07}). DB8's distance has been determined spectrophotometrically by \cite{Messineo07} to be 2.65 ${\pm}$ 0.4 kpc. We adopt this same distance for DB7. If the two clusters are associated with the same H II complex, DB7 may have formed first and triggered the collapse of DB8 (\citealt{Messineo07}).

\subsection{DBCL23}

The most recent detailed study of DBCL23 was done by \cite{Comeron05}, who estimated an upper limit of 3.5 kpc for its distance. The spectrum of the nebula around this source indicates a temperature of 40,000 - 45,000 K (corresponding to an O5 - O7 star) for the ionizing stellar spectrum. However, \cite{Comeron05} discovered a significant discrepancy between the ionizing flux supplied by a mid O-type star and that inferred from the optical and radio parameters. It is possible that DBCL23 is ionized by an external O-type star, although the ionizing source was not identified. Their results confirm that DBCL23 is not a massive star forming site.

\section{OBSERVATIONS AND IMAGE PROCESSING}

Radio observations were made at 8.5 and 4.9 GHz with the Very Large Array (VLA) telescope of the National Radio Astronomy Observatory (NRAO) \footnotemark\footnotetext{The National Radio Astronomy Observatory is a facility of the National Science Foundation operated under cooperative agreement by Associated Universities, Inc.} in August 2008 through observing program AL723. The VLA was in its D array configuration allowing for resolutions of 5-20\arcsec. At both frequencies, J1331+305 was used as a flux density calibrator. Standard procedures for calibration, editing and imaging were carried out using the Astronomical Image Processing Software (AIPS) of the NRAO. Table 2 summarizes the image parameters for the different sources. All images were made with robust = 0 weighting.

Images of all six targets were taken in 3.6, 4.5, 5.8, and 8.0 $\mu$m bands using the Infrared Array Camera (IRAC) onboard the {\it Spitzer Space Telescope} (Program ID 30734, PI Figer). The 5 position Gaussian dithering pattern was used, and the resulting images in each wavelength cover a 5\arcmin~by 5\arcmin~area, with a pixel size of 1.2\arcsec. The total integration time was 471 s in each of the four bands. The {\it Spitzer} MOPEX tool was applied to the basic calibrated data to create mosaicked images of each band. An IDL adapted version of DAOPHOT was used to perform photometry on each band, and the results were cross-correlated. If a source was not present in one or more of the 4.5, 5.8, or 8.0 $\mu$m bands, we reported a null result for the specific band(s), rather than remove the source entirely.

\section{RESULTS}

\begin{figure}
\epsscale{1.0}
\plotone{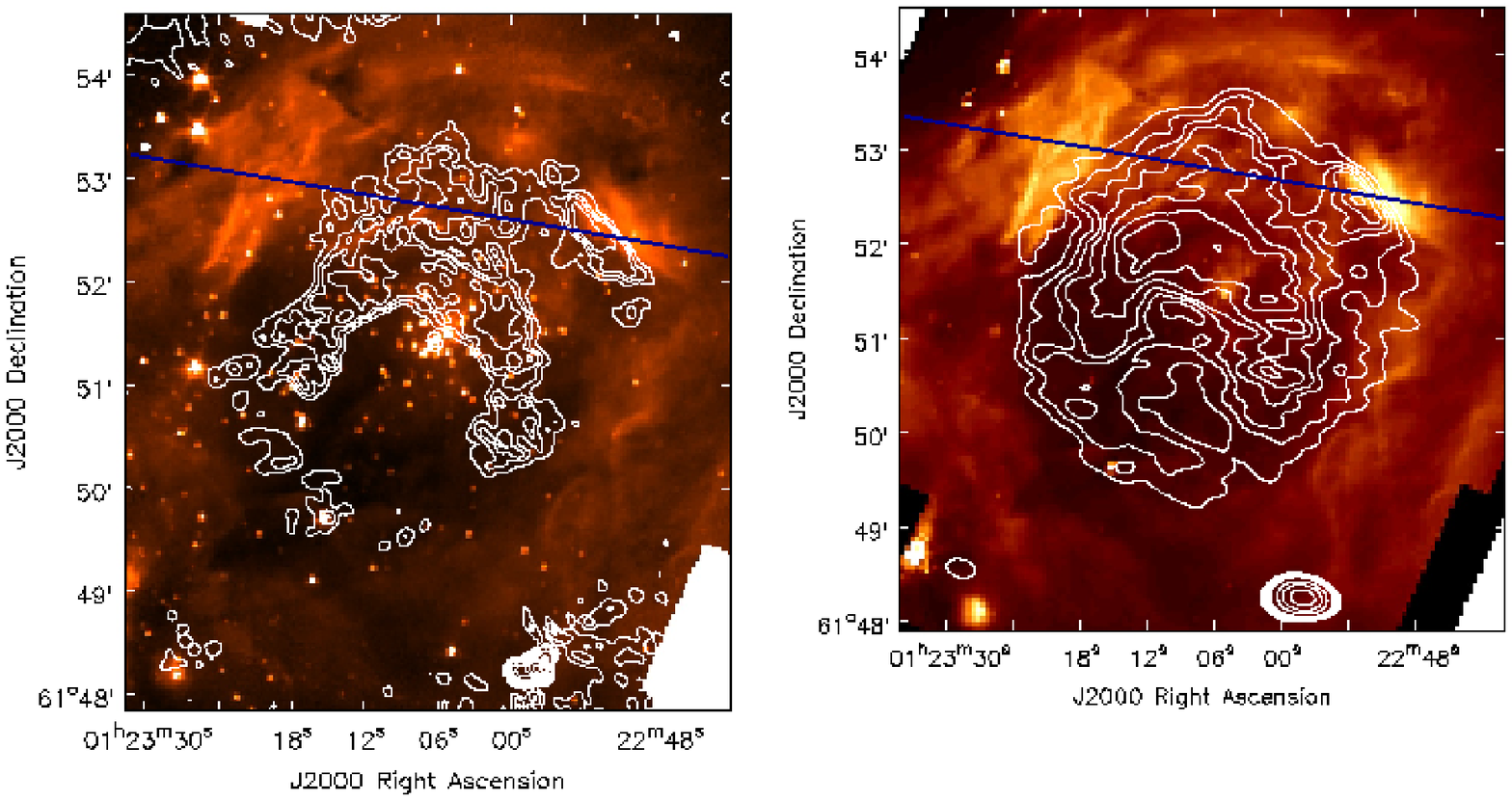}
\caption{BD52: The left image shows radio contours at 8.5 GHz corresponding to levels of 3, 6, 9, 15, 30, 60, 120, 200, 300, 450 times rms noise of 3.65$\times$10$^{-5}$ Jy beam$^{-1}$ overlaid on color scale representing {\it Spitzer} IRAC data at 3.6 $\mu$m. The right image shows radio countours at 4.9 GHz corresponding to levels of 3, 5, 7, 9, 12, 15, 18, 50, 100, 200 times rms noise of 3.13$\times$10$^{-4}$ Jy beam$^{-1}$ overlaid on color scale representing {\it Spitzer} IRAC data at 8 $\mu$m. The blue lines represent the slice taken through the shell seen in Figures \ref{fig9} and \ref{fig10}. The synthesized beam of the radio image is shown in the lower left corner of each plot.
\label{fig1}}
\end{figure}

\begin{figure}
\epsscale{1.0}
\plotone{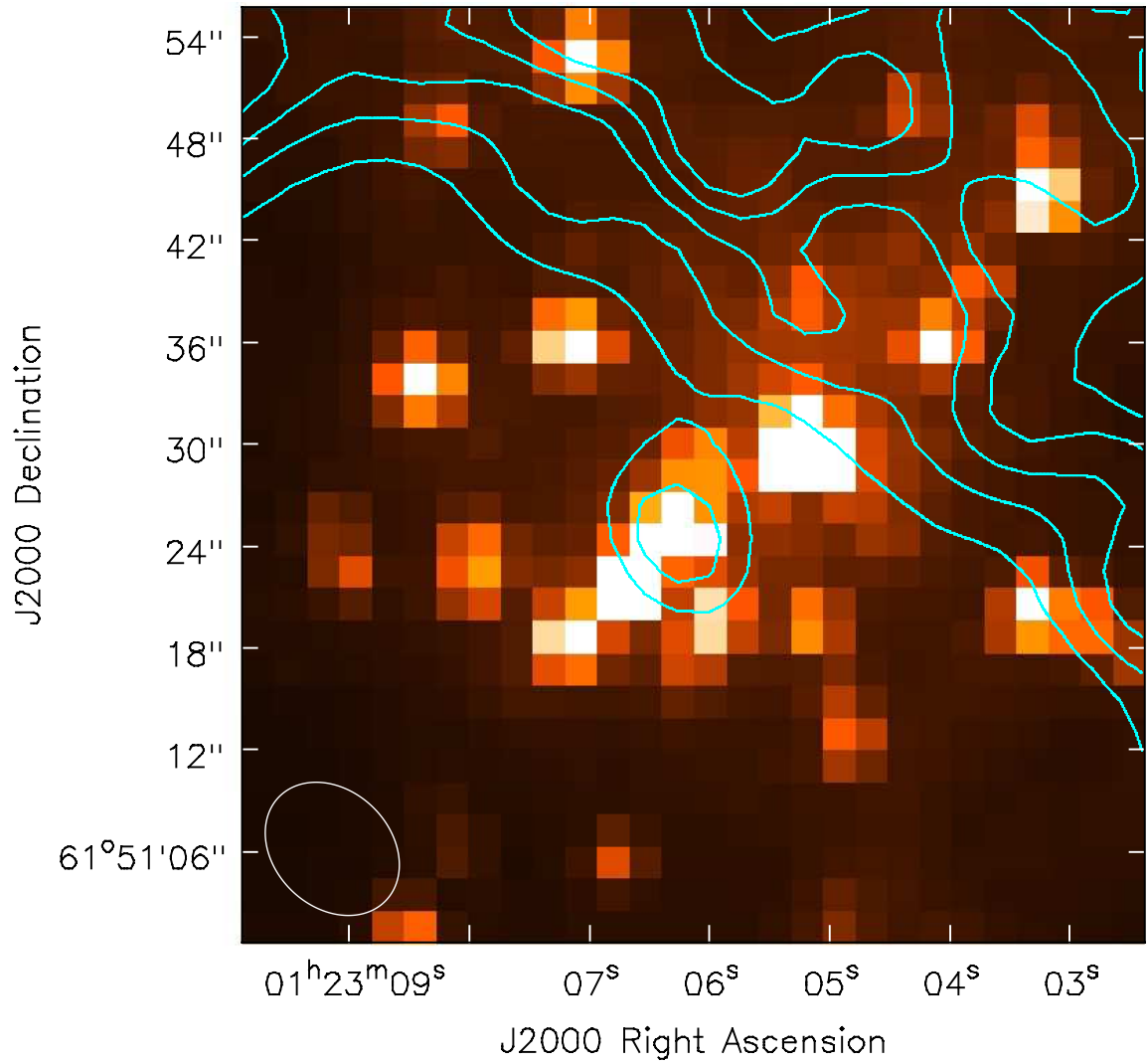}
\caption{BD52 center with radio contours at 8.5 GHz corresponding to levels of 3, 6, 9, 12, 15 times rms noise of 3.65$\times$10$^{-5}$ Jy beam$^{-1}$ overlaid on color scale representing {\it Spitzer} IRAC data at 3.6 $\mu$m. The synthesized beam of the radio image is shown in the lower left corner of the plot.
\label{fig2}}
\end{figure}

\begin{figure}
\epsscale{1.0}
\plotone{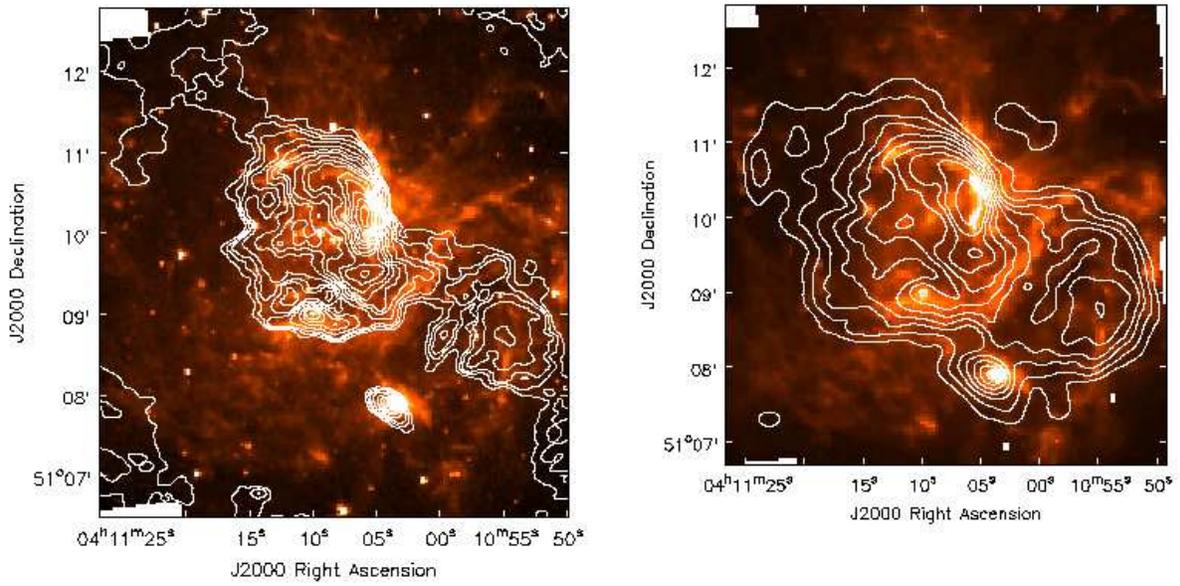}
\caption{BD65: The left image shows radio contours at 8.5 GHz corresponding to levels of 3, 6, 9, 15, 25, 35, 45, 60, 75, 90, 120, 150, 180, 220, 260 times rms noise of 1.56$\times$10$^{-4}$ Jy beam$^{-1}$ overlaid on color scale representing {\it Spitzer} IRAC data at 3.6 $\mu$m. The right image shows radio contours at 4.9 GHz corresponding to levels of 3, 8, 15, 22, 32, 45, 60, 80, 100, 150 times rms noise of 5.52$\times$10$^{-4}$ Jy beam$^{-1}$ overlaid on color scale representing {\it Spitzer} IRAC data at 8 $\mu$m. The synthesized beam of the radio image is shown in the lower left corner of each plot.
\label{fig3}}
\end{figure}

\begin{figure}
\epsscale{1.0}
\plotone{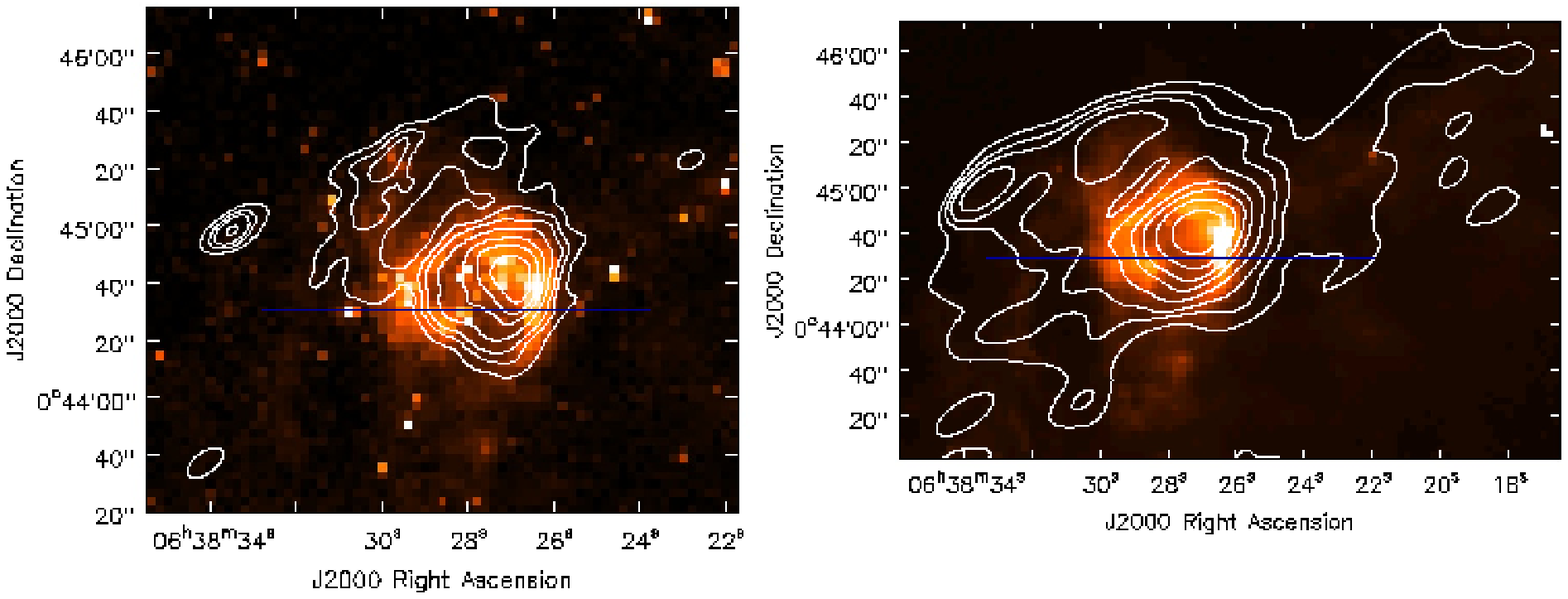}
\caption{BD84: The left image shows radio contours at 8.5 GHz corresponding to levels of 3, 6, 9, 15, 25, 35, 50, 70 times rms noise of 4.25$\times$10$^{-5}$ Jy beam$^{-1}$ overlaid on color scale representing {\it Spitzer} IRAC data at 3.6 $\mu$m. The right image shows radio contours at 4.9 GHz corresponding to levels of 3, 6, 9, 15, 25, 35, 50, 70 times rms noise of 6.75$\times$10$^{-5}$ Jy beam$^{-1}$ overlaid on color scale representing {\it Spitzer} IRAC data at 8 ${\mu}$m. The blue lines represent the slice taken through the shell seen in Figures \ref{fig13} and \ref{fig14}. The synthesized beam of the radio image is shown in the lower left corner of each plot.
\label{fig4}}
\end{figure}

\begin{figure}
\epsscale{1.0}
\plotone{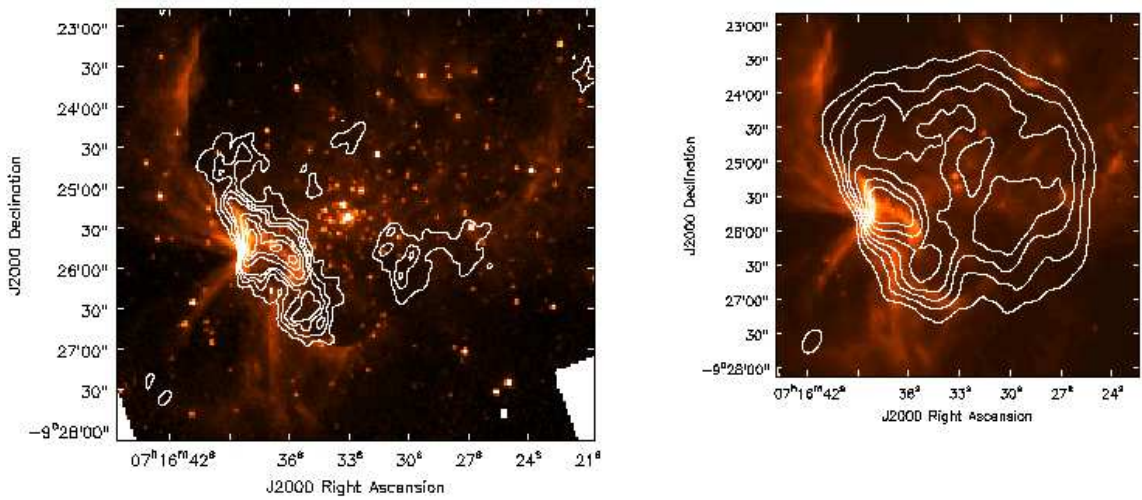}
\caption{BD95: The left image shows radio contours at 8.5 GHz corresponding to levels of 3, 5, 7, 9, 14, 18, 22 times rms noise of 5.91$\times$10$^{-5}$ Jy beam$^{-1}$ overlaid on color scale representing {\it Spitzer} IRAC data at 3.6 ${\mu}$m. The right image shows radio contours at 4.9 GHz corresponding to levels of 3, 5, 7, 9, 12, 16, 20 times rms noise of 2.81$\times$10$^{-4}$ Jy beam$^{-1}$ overlaid on color scale representing {\it Spitzer} IRAC data at 8 $\mu$m. The synthesized beam of the radio image is shown in the lower left corner of each plot.
\label{fig5}}
\end{figure}

\begin{figure}
\epsscale{1.0}
\plotone{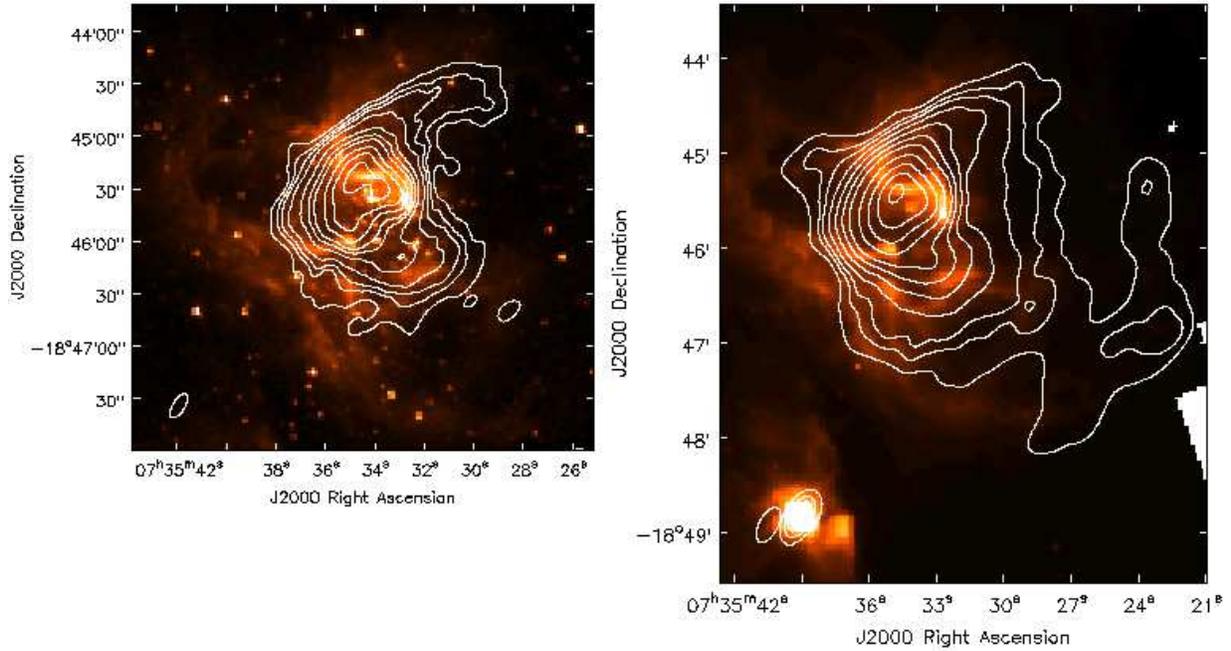}
\caption{DB7: The left image shows radio contours at 8.5 GHz corresponding to levels of 3, 6, 9, 15, 30, 45, 60, 90, 120, 160 times rms noise of 1.06$\times$10$^{-4}$ Jy beam$^{-1}$ overlaid on color scale representing {\it Spitzer} IRAC data at 3.6 $\mu$m. The right image shows radio contours at 4.9 GHz corresponding to levels of 3, 6, 9, 15, 25, 40, 60, 100, 150, 200, 250 times rms noise of 1.94$\times$10$^{-4}$ Jy beam$^{-1}$ overlaid on color scale representing {\it Spitzer} IRAC data at 8 ${\mu}$m. The synthesized beam of the radio image is shown in the lower left corner of each plot.
\label{fig6}}
\end{figure}

\begin{figure}
\epsscale{1.0}
\plotone{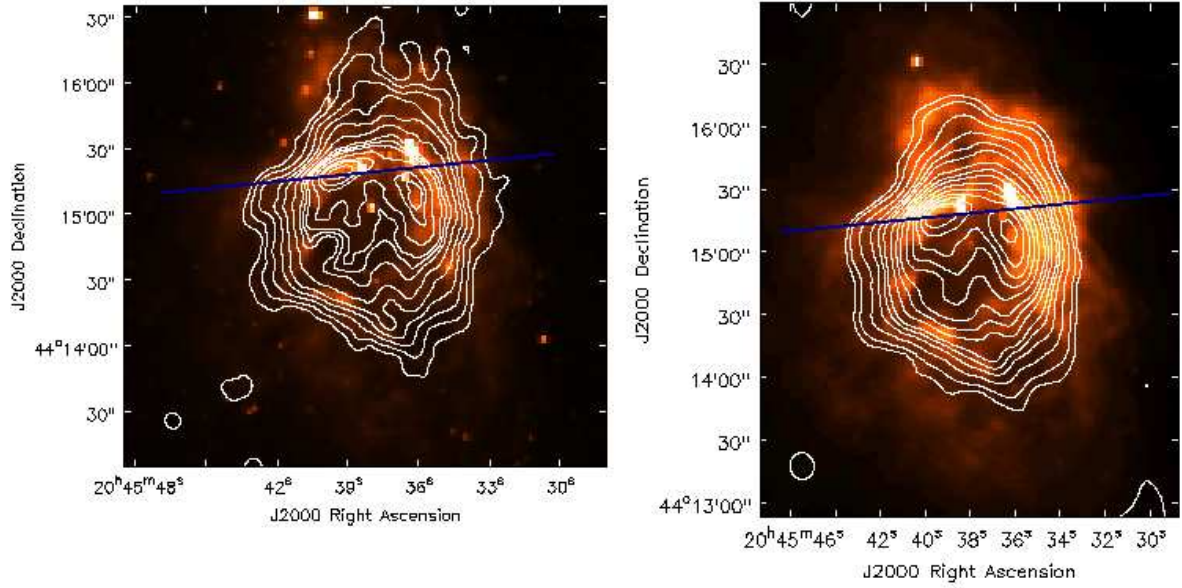}
\caption{DBCL23: The left image shows radio contours at 8.5 GHz corresponding to levels of 3, 8, 18, 32, 50, 80, 120, 145, 170, 225, 290, 325, 375 times rms noise of 1.34$\times$10$^{-4}$ Jy beam$^{-1}$ overlaid on color scale representing {\it Spitzer} IRAC data at 3.6 ${\mu}$m. The right image shows radio contours at 4.9 GHz corresponding to levels of 3, 6, 15, 30, 45, 60, 90, 120, 150, 200, 250, 300, 350, 400 times rms noise of 2.95$\times$10$^{-4}$ Jy beam$^{-1}$ overlaid on color scale representing {\it Spitzer} IRAC data at 5.8 ${\mu}$m. The blue lines represent the slice taken through the shell seen in Figures \ref{fig11} and \ref{fig12}. The synthesized beam of the radio image is shown in the lower left corner of each plot.
\label{fig7}}
\end{figure}

In the following sections the morphology of the clusters and their surroundings in both radio and infrared images is discussed. The correlation between the observed radio and infrared emission in Figures \ref{fig1}-\ref{fig7} is also examined. The wavelengths at which the observations were taken provide information about the type of emission produced by the stellar clusters. Ionized hydrogen in the clusters emits thermal free-free emission, which can be traced at radio frequencies. Emission from the thermal dust continuum is seen at infrared frequencies. The IRAC bands at 3.6, 5.8, and 8 $\mu$m contain polycyclic aromatic hydrocarbon (PAH) features. PAH emission typically arises in PDRs and can be used to delineate the edge of the PDR. The 3.6 $\mu$m band is dominated by stellar emission, and the 8 $\mu$m band contains primarily diffuse continuum and PAH emission. These two bands were used for the colorscale seen in Figures \ref{fig1}-\ref{fig7} with contours overlaid representing the radio emission at 8.5 and 4.9 GHz. Comparing the distribution of radio and infrared emission provides us with knowledge about the morphology and emission mechanisms of the stellar clusters and their H II regions. 

\subsection{Radio and Infrared Images}

\subsubsection{BD52}

Figure \ref{fig1} shows BD52 with contours of radio emission (at 8.5 GHz (left) and 4.9 GHz (right)) overlaid on 3.6 and 8 $\mu$m images. The radio emission associated with BD52 has an extended morphology. The 8.5 GHz radio contours in the left image show a curved, partial-shell structure of approximately 1\arcmin~(thickness) x 3\arcmin~(diameter) (0.4 pc x 1.3 pc at d = 1.44 kpc) in size, whereas the 4.9 GHz radio emission shows a complete shell-like structure as seen in the image on the right. Given the distribution of radio emission at 4.9 GHz, it appears that there must be missing flux at 8.5 GHz. Indeed, the largest angular size of the VLA interferometer at 8.5 GHz in D-array is $\sim$3\arcmin. This means that extended structures larger than this angular size (3\arcmin) will not be detectable at this frequency. The 4.9 GHz emission associated with BD52 shows that the source has a diameter of about 4\arcmin~(1.7 pc at d = 1.44 kpc), therefore, flux is missing in the 8.5 GHz image. Two components are visible in the 3.6 and 8 $\mu$m emission: the stellar sources and the diffuse emission that appears to be part of a larger infrared shell extending beyond the 4.9 GHz radio shell. The emission is more apparent at 8 $\mu$m shown in the right image, and to the north looks more complete than at 3.6 $\mu$m. The stellar cluster is visible in the 3.6 $\mu$m image on the left. An investigation of the stellar components of BD52 by Trombley et al. (in prep.) suggests a cluster mass of 1300 M$_{\sun}$ and an age of 1 Myr. They conclude that several main sequence early B stars are producing the ionizing flux which powers the H II region.

In the SW corner of the image is a bright radio point source (at approximately RA, DEC (J2000): 01 23 00, 61 48 15). This radio point source has no counterpart in the infrared and is known to be a bright non-thermal extragalactic radio source (\citealt{S-B86}). We calculate the spectral index of this compact source to be -2.6$\pm$0.07, which is in agreement with the non-thermal nature of the source. 

Figure \ref{fig2} shows a small area from the center of BD52 (approximately 50\arcsec~by 50\arcsec, corresponding to 0.35 x 0.35 pc at d = 1.44 kpc) with 8.5 GHz radio contours overlaid on the 3.6 $\mu$m image. In the center of the stellar cluster, there is a point-like radio source that appears to be coincident with an infrared source. The 8.5 GHz flux is 0.3$\pm$0.1 mJy. This association is suggestive and one interpretation is that the radio emission arises from a stellar wind in one or more of the members of the stellar cluster. Presuming the radio emission associated with the stellar source is due to the massive stellar wind, it is possible to calculate the mass-loss rate using Equation (1) of \cite{Lang01b}, based on \cite{P-F75}. Assuming a terminal wind velocity of  $\sim$1000 km s$^{-1}$ and a distance of 1.44 kpc, the mass-loss rate is approximately 0.35$\times$10$^{-5}$ M$_\sun$ yr$^{-1}$. This mass-loss rate is somewhat higher than expected for an early B-type main sequence star (\citealt{Sternberg03}). This particular radio source also has counterparts in the near-IR (HST/NIC3), mid-IR (Spitzer/IRAC), and X-ray (CXO) (Trombley et al. (in prep)).

\subsubsection{BD65}

Figure \ref{fig3} shows BD65 with contours of radio emission (at 8.5 GHz (left) and 4.9 GHz (right)) overlaid on 3.6 and 8 $\mu$m images. The 3.6 and 8 $\mu$m infrared emission associated with BD65 is primarily diffuse with denser cores of bright emission that form a shell-like structure around the central cluster located at  RA, DEC (J2000): 04 11 08, 51 10 00. The shell of infrared emission is outlined by the 8.5 GHz radio emission with a diameter of $\sim$2\arcmin~(5.7 pc at d = 9.8 kpc) shown in contours in the left image. This emission also extends to the NE and to the SW of the central stellar cluster. The 4.9 GHz radio emission in the right image is more extended than the 8.5 GHz emission. The morphology defined by this radio emission differs from that seen in the 3.6 and 8 $\mu$m emission. The two brightest cores of infrared emission (e.g. RA, DEC (J2000): 04 11 05, 51 10 00; 04 11 05, 51 10 30) are also noticeable in the radio emission, but are not resolved into two distinct components. There is another notable region south of the central cluster at RA, DEC (J2000): 04 11 04, 51 08 00 that is visible at both radio and infrared frequencies. This compact region of bright emission approximately 0.5\arcmin~(1.4 pc at d = 9.8 kpc) in size is not associated with the extended emission at 8.5 GHz, but is part of the extended emission at 4.9 GHz.

\subsubsection{BD84}

Figure \ref{fig4} shows BD84 with contours of radio emission (at 8.5 GHz (left) and 4.9 GHz (right)) overlaid on 3.6 and 8 $\mu$m images. The radio emission at both frequencies for BD84 shows a symmetric distribution of contours around the central cluster about 1\arcmin~(2.3 pc at d = 8.06 kpc) in size with surrounding diffuse emission. The emission at 3.6 and 8 $\mu$m shows a shell-like structure with bright edges, whereas the radio emission is more complete with a single central peak in emission. The 4.9 GHz radio emission in the image on the right contains a large amount of diffuse emission that extends well past the infrared emission with an approximate size of 3\arcmin~x 2\arcmin~(7.0 x 4.7 pc at d = 8.06 kpc). The 8.5 GHz radio emission in the left image only extends NE of the cluster. The peak for the infrared emission lies west of the central cluster at about RA, DEC (J2000): 06 38 26, 0 44 30. The peak in radio emission, however, appears closer to the center of the cluster at RA, DEC (J2000): 06 38 27, 0 44 45. The radio source at RA, DEC (J2000): 06 38 34, 0 45 00 appears distinct in the 8.5 GHz radio emission, but is part of the extended emission at 4.9 GHz. 

\subsubsection{BD95}

BD95 is shown in Figure \ref{fig5} with contours of radio emission (at 8.5 GHz (left) and 4.9 GHz (right)) overlaid on 3.6 and 8 $\mu$m images. The 4.9 GHz radio contours show a complete shell with a diameter of $\sim$5\arcmin~(4.7 pc at d = 3.24 kpc). There is corresponding 3.6 and 8 $\mu$m infrared emission only along the eastern side where the brightest infrared emission can be seen. The 8.5 GHz contours shown in the left image appear drastically different with emission only around the brightest infrared emission. As with BD52, it appears that flux is missing at 8.5 GHz, most likely due to the aforementioned largest angular size limitations of the VLA. The structure of the 3.6 and 8 $\mu$m emission is complex in comparison to the simple shell-like 4.9 GHz emission. The infrared images consist of a bright region of emission at RA, DEC (J2000): 07 16 39, -09 25 45 with diffuse arm-like features emanating out to the NE and SW. This peak emission is host to an embedded cluster (see \S2.4). The cluster containing the ionizing source is visible at 3.6 $\mu$m in the left image and is located NW of the cluster at RA, DEC (J2000): 07 16 33, -09 25 15. 

\subsubsection{DB7}

The radio emission associated with DB7 is shown in contours (at 8.5 GHz (left) and 4.9 GHz (right)) overlaid on 3.6 and 8 $\mu$m images in Figure \ref{fig6}. The infrared emission in DB7 has a different morphology than the radio emission. The 8.5 and 4.9 GHz radio emission contains a denser circular region of contours around the cluster, and some surrounding diffuse emission. The densest radio contours which encompass the central cluster (seen at 3.6 $\mu$m) have a diameter of $\sim$2\arcmin~(1.5 pc at d = 2.65 kpc). The 4.9 and 8.5 GHz radio emission have similar morphology, although the 4.9 GHz emission in the right image extends farther west. The 3.6 and 8 $\mu$m infrared morphology resembles that of BD95 with areas of bright emission and diffuse arm-like structures. One of these structures which arcs out to the NW has corresponding radio emission, most noticeable at 8.5 GHz in the left image. Other than this structure, the radio and infrared are not well correlated. The emission peaks are offset with the infrared peak occurring at RA, DEC (J2000): 07 35 33, -18 45 30, and the radio peak occurring at RA, DEC (J2000): 07 35 35, -18 45 30. The associated source, DB8, can be seen in the right image at RA, DEC (J2000): 07 35 39, -18 48 45. 

\subsubsection{DBCL23}

Figure \ref{fig7} shows DBCL23 with contours of radio emission (at 8.5 GHz (left) and 4.9 GHz (right)) overlaid on 3.6 and 5.8 $\mu$m images. The radio contours at both frequencies are very similar, and both are well correlated to the infrared emission. There is only a slight difference in morphology between the infrared and radio with the radio emission being filled-in, and the infrared emission more shell-like with a central cavity. The contours of 4.9 GHz emission shown in the image on the right give a diameter of approximately 2\arcmin~(2 pc at d = 3.5 kpc) for DBCL23. The diffuse 5.8 $\mu$m emission seen in the right image extends farther than the 4.9 and 8.5 GHz radio contours. There are double emission peaks separated by about 30\arcsec~visible in the radio and infrared emission. The location of these peaks in the radio emission is offset from the peaks in the infrared emission.

\subsection{Relationship Between Radio and Infrared Emission}

Examination of Figures \ref{fig1}-\ref{fig7} reveal some overall trends for the ISM surrounding the candidate clusters: (1) In all but two sources, the infrared emission (3.6 and 8 $\mu$m) extends beyond the associated radio emission. The two exceptions to this are BD84 and DB7, for which the 4.9 GHz emission extends farther than the infrared emission; (2) The morphology of the infrared emission for the cluster sources BD52, BD84, and DBCL23 is shell-like with a cavity and bright emission along the edges. The radio emission in BD52, BD84, and DBCL23 is also more symmetric and roughly follows the infrared emission. The shell-like morphology at both radio and infrared wavelengths allows for a more in-depth analysis to be carried out for these three sources (\S5.1); (3) On the other hand, BD65, BD95, and DB7 are more complex and less symmetric in morphology at radio and infrared wavelengths; and (4) the peak intensity seen in the radio emission is offset from the peak in the infrared emission. This trend is also noticeable in the slice analysis performed on the three cluster candidates BD52, DBCL23, and BD84 (see \S5.1). 

\subsection{Radio Continuum Properties of H II Regions}

\subsubsection{Calculations}

\begin{figure}
\epsscale{0.8}
\plotone{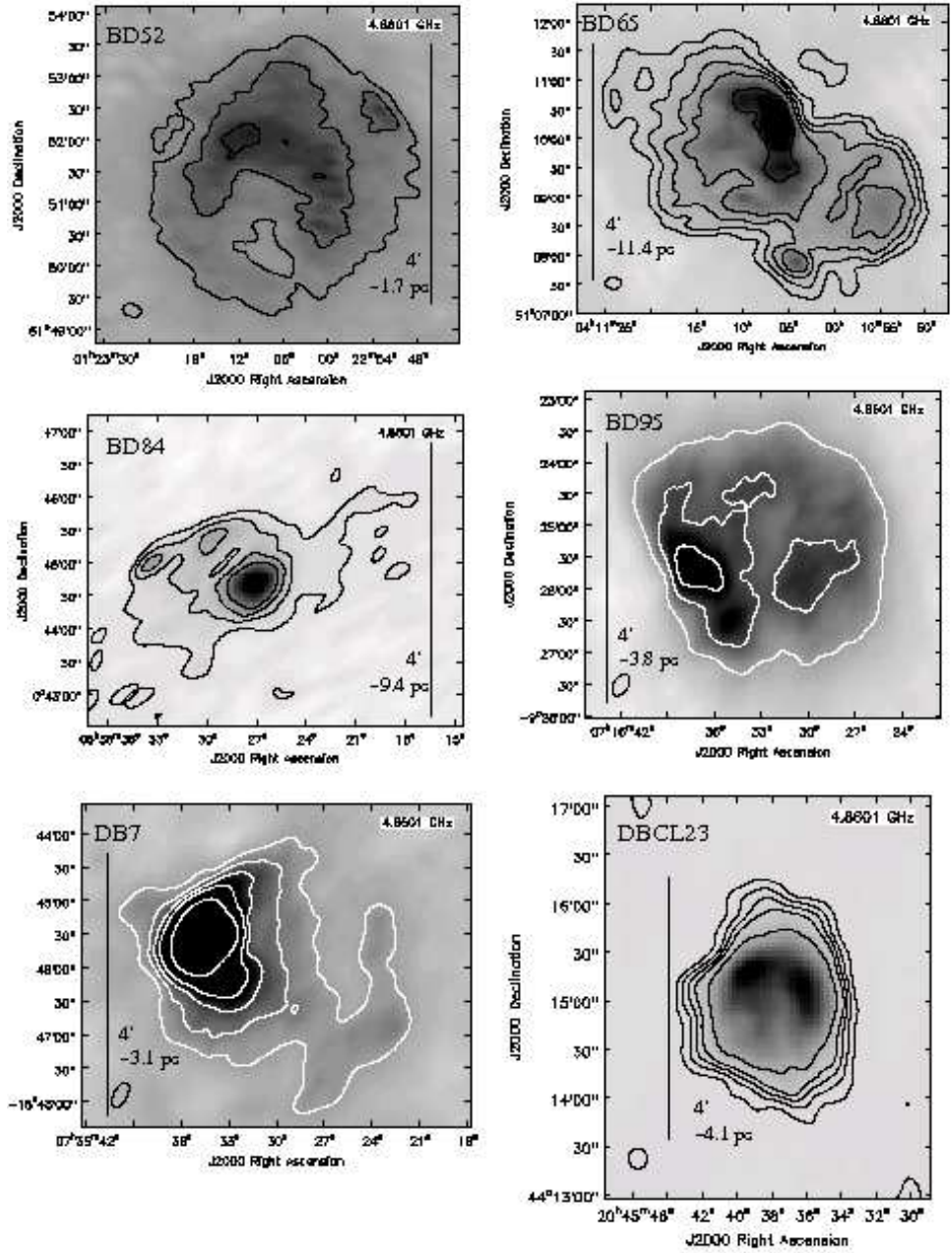}
\caption{Images show the 4.9 GHz radio continuum with contours at 3, 9, 18, 32, 80 times rms noise in each source. Flux densities were integrated for each source using the 3$\sigma$ contour. A 4' bar with the corresponding linear size at the source's distance has been included to demonstrate the relative size scale.
\label{fig8}}
\end{figure}

We can estimate the physical properties of the radio continuum emission by examining the flux density of each source. Flux densities have been measured within the contour representing 3 times the rms level at 4.9 GHz as seen in Figure \ref{fig8}. Approximate angular sizes have also been determined from the full width at zero intensity of slices taken across each source. The 3$\sigma$ contour at 4.9 GHz was used to delineate the edge of the sources. The integrated flux density and angular size values are reported in Table \ref{tbl-3}.

Based on the radio continuum parameters, physical properties of the ionized gas associated with the candidate clusters have been derived using the formulation of \cite{M-H67}. In using these formulae, a uniform density, spherical, ionization bounded H II region with an electron temperature of 10,000 K and an abundance of ionized helium to hydrogen of 0.1 is assumed. Table \ref{tbl-4} presents the derived quantities from the radio continuum parameters shown in Table \ref{tbl-3}. Due to flux missing from a couple of the sources in the 8.5 GHz observations, an accurate measure of the flux density could not be obtained, thereby preventing a useful analysis at this frequency. Using the formulae, we derive a linear radius of the equivalent sphere for the H II region (r), electron density (n$_e$), emission measure (EM), total mass of ionized gas (M$_{\sun}$), and the Lyman continuum photon flux (N$_{Lyc}$). The Lyman continuum photon flux is the rate of photons emitted capable of ionizing the surrounding hydrogen. This value provides a measure of the cluster's power by quantifying its ability to impact the surrounding gas. 

The authors acknowledge that the sources presented in this study are not well described by a uniform density sphere. However, past studies (most notably the work done by \citealt{Balser95}) have shown H II region properties based on radio continuum emission to have relatively limited model dependence. \cite{Balser95} compared three models to describe the radio continuum structure of 11 Galactic H II regions. Physical parameters were derived from the models using different geometries (homogeneous sphere and a spherical Gaussian) as well as by using peak properties in the continuum intensity maps. Comparison between the three models showed no systematic differences in the derived electron density and emission measure, and resulted in an overall scatter of 15$\%$ between modeled parameters. In addition, the ionization properties which are of greatest interest in this analysis are a function of frequency, electron temperature, distance, and flux density, all of which are model independent. The main challenge lies in accurately measuring the flux density. \cite{Balser95} discovered there could be differences in measured flux density of up to $\sim$50$\%$ for extended sources, which would result in about a 50$\%$ difference in derived ionizing flux. However, the motivation of this work is an order of magnitude comparison of ionizing photon fluxes between the sources in this study and known massive clusters in the Galaxy. Therefore, the results of this analysis are adequate for the primary goal of this study. It should be emphasized that the physical quantities reported in Table \ref{tbl-4} derived from the radio continuum parameters are meant to be used for relative comparison.


The last column in Table \ref{tbl-4} gives an estimate of the spectral type for each cluster based on our calculated Lyman continuum photon fluxes and model stellar atmospheres derived by \cite{Smith02}. No correction for extinction was made for these estimates. The stellar spectral types given are the types that would be assigned to a single ionizing source which produces the observed Lyman continuum flux. It is recognized that the ionizing photon rate calculated for the stellar clusters most likely arises from more than one cluster member, and not a single ionizing source of the estimated spectral type. However, the estimate allows for a comparison between the results of this analysis and those determined in previous studies. In general, there was close agreement between the estimated spectral types indicating these results are consistent with past observations (e.g. BD95; \citealt{Samal07,Yun08}).

\subsubsection{Properties}

The physical parameters derived in Table \ref{tbl-4} show variation in physical characteristics of the H II regions associated with the cluster candidate sources. The equivalent radii for the H II regions range between $\sim$1 and 9 pc, with the majority of sources having linear sizes of $\sim$1-3 pc. The electron densities range from 10 cm$^{-3}$ up to 140 cm$^{-3}$ with an average of $\sim$48 cm$^{-3}$. The EM differs for all sources ranging between $\sim$10$^3$ pc cm$^{-6}$ and 6$\times$10$^4$ pc cm$^{-6}$. The total mass of ionzed gas associated with the H II regions varies widely from 7 to 2800 M$_{\sun}$. The cluster source BD52 has the smallest ionized mass of 7.4 M$_{\sun}$. BD84, BD95, DB7, and DBCL23 have intermediate masses averaging around $\sim$50 M$_{\sun}$, whereas BD65 contains a remarkable 2800 M$_{\sun}$ of ionized gas. The Lyman continuum photon fluxes spread over two orders of magnitude from $\sim$10$^{47}$ s$^{-1}$ to 10$^{49}$ s$^{-1}$ with an average of $\sim$6$\times$10$^{48}$ s$^{-1}$. This spread in ionizing photon rates corresponds to a spread in stellar spectral type from early B-type to mid O-type. The majority of the candidate sources have Lyman continuum photon fluxes that would be produced by single B0 type star.

In this candidate cluster comparison, BD65 stands out as having greater values for most derived parameters. It clearly has the largest linear radius of 9.1 pc, much greater than the other clusters' average of $\sim$2 pc. Similarly, BD65 has an H II mass that is nearly 70 times greater than the average of the other clusters' ionized mass. BD65 has the greatest Lyman continuum photon flux of 3.5$\times$10$^{49}$ s$^{-1}$, corresponding to an ionizing source of spectral type O3V. The physical paramters of the H II regions associated with these clusters depend on the distance. Therefore, the discrepancy in distance estimates to BD65 implies the physical properties derived would be less extreme if a smaller distance was adopted. For example, at a distance of 4.9 kpc (\citealt{Bica03}) BD65 has a linear radius of 4.5 pc, an ionized mass of 490 M$_{\sun}$, and a Lyman continuum photon flux of 8.8$\times$10$^{48}$ s$^{-1}$. Even at this smaller distance BD65 has the largest linear size, ionized mass, and ionizing photon flux of the candidate sources. Although small in size and mass as compared to BD65, DBCL23 has the largest electron density and emission measure. DBCL23 also has the second highest ionizing photon flux (2.7$\times$10$^{48}$ s$^{-1}$), which is about a magnitude less than that of BD65.

\section{DISCUSSION}

\subsection{Radio and Infrared Comparisons}

\cite{Watson08} performed an analysis on infrared dust bubbles using radio and mid-infrared data for bubbles in the Galactic plane which led to a generalized picture of infrared bubbles as ionized gas with hot dust surrounded by a PDR containing interstellar gas, PAHs, and dust. The inner ionized gas is traced by radio free-free emission, and the thermal dust continuum can be traced with infrared emission, which often lies outside of the region of radio emission. The surrounding PDR is dominated by PAH emission features which can be seen in IRAC bands at 3.6, 5.8, and 8 $\mu$m. Here, we carry out a similar analysis using our radio and infrared data for several of the sources that appear to be most symmetric: BD52, DBCL23, and BD84.

\subsubsection{BD52}

\begin{figure}
\epsscale{1.0}
\plotone{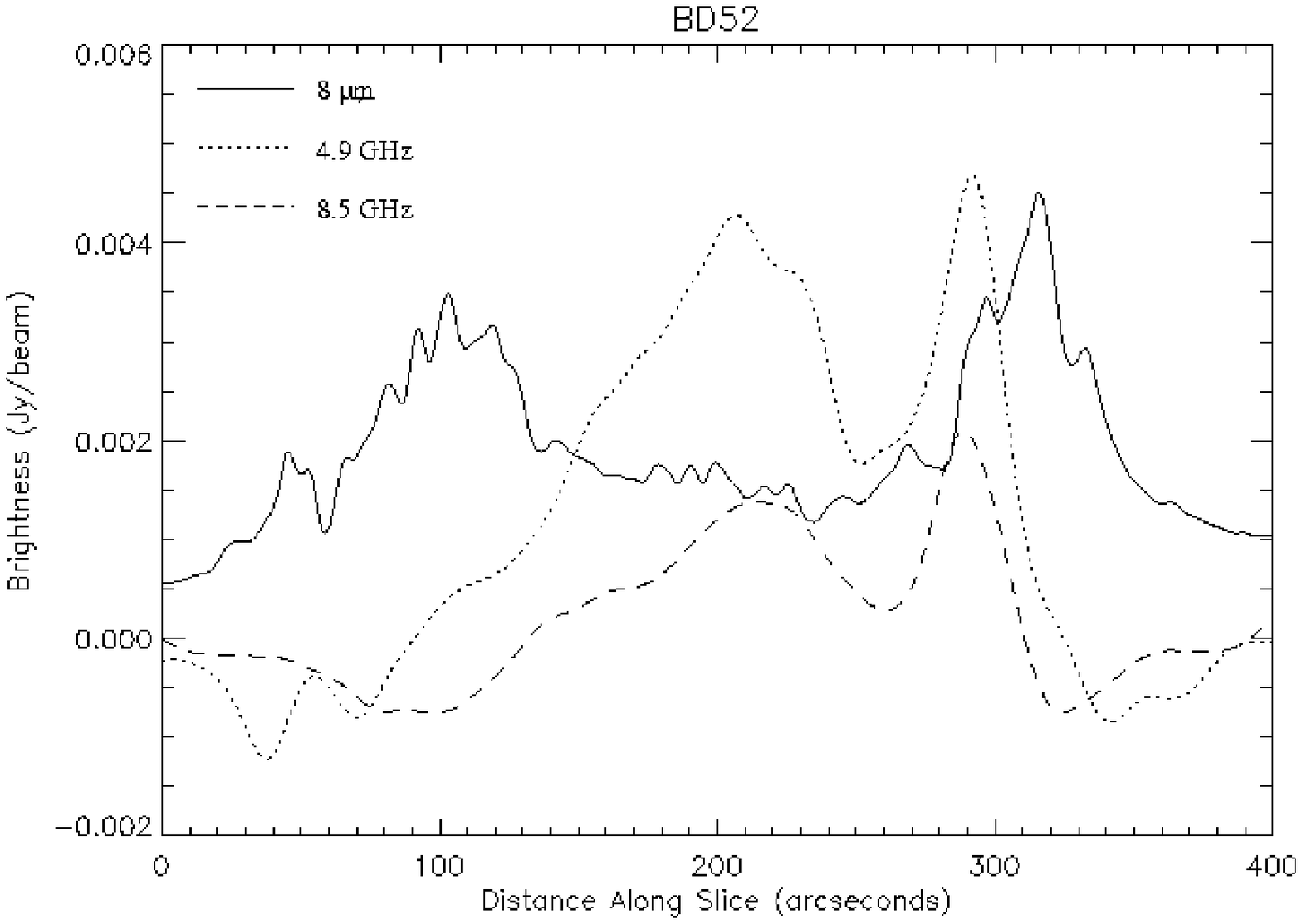}
\caption{BD52 slice near DEC = 61\degr53\arcmin~at 8 $\mu$m (solid line), 4.9 GHz (dotted line), and 8.5 GHz (dashed line). The position of the slice is shown exactly in Figure \ref{fig1}. The 8 $\mu$m intensity has been scaled down by a factor of 10$^5$. 
\label{fig9}}
\end{figure}

\begin{figure}
\epsscale{1.0}
\plotone{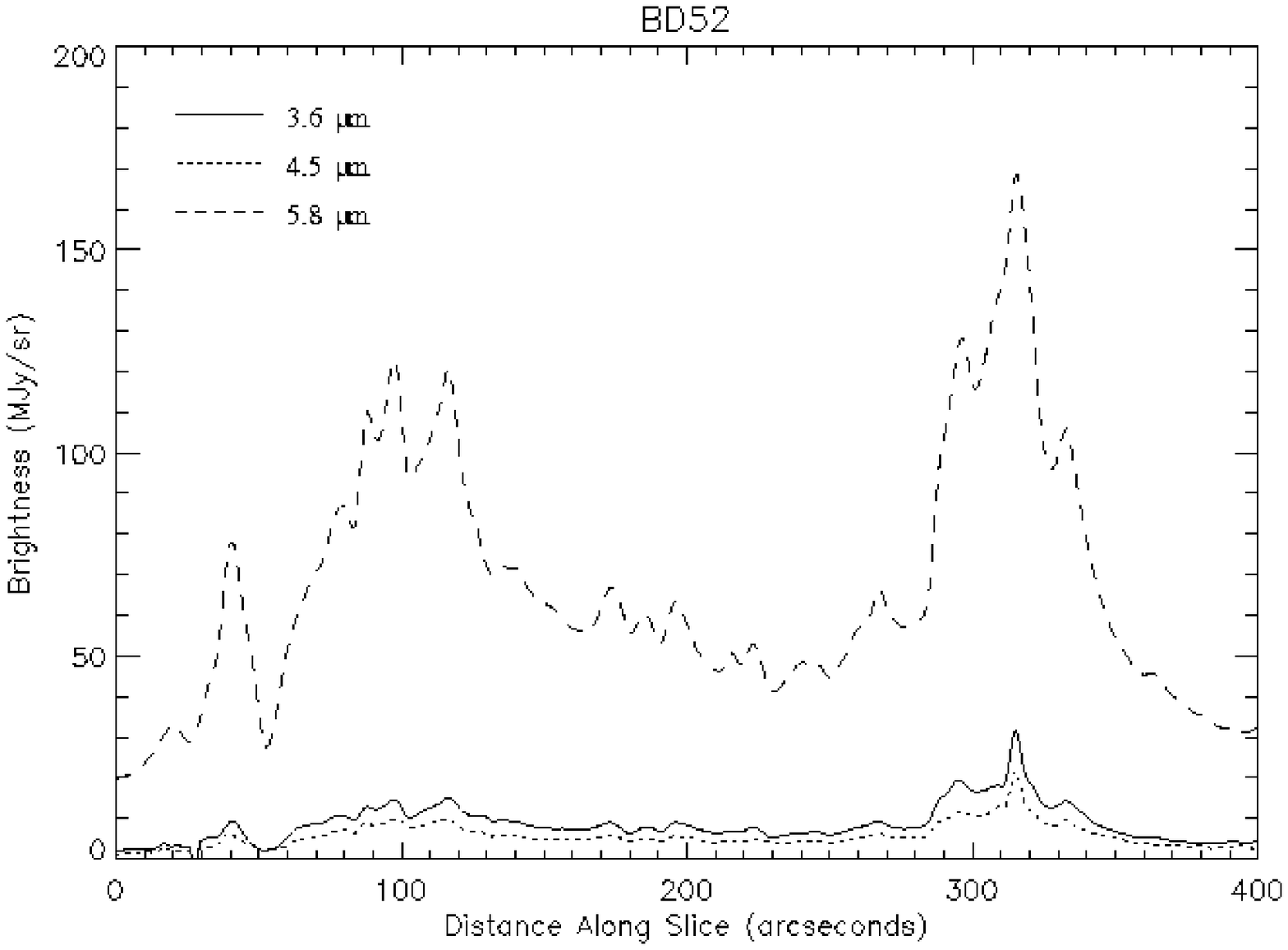}
\caption{BD52 slice near DEC = 61\degr53\arcmin~at 3.6 (solid line), 4.5 (dotted line), and 5.8 $\mu$m (dashed line). The position of the slice is shown exactly in Figure \ref{fig1}.
\label{fig10}}
\end{figure}

Figure \ref{fig9} shows a slice through the BD52 shell at 4.9 GHz, 8.5 GHz (convolved to match the 4.9 GHz) and 8 $\mu$m. The location of this slice is shown by the blue line in Figure \ref{fig1}. The slight difference in flux at 8.5 GHz compared to 4.9 GHz is most likely due to missing flux, as previously discussed in \S4.1.1. Figure \ref{fig9} shows that the radio emission peaks in two places near 200\arcsec~and 300\arcsec~along the slice. The 8 $\mu$m emission exhibits the same double peaked structure, but peaks at $\sim$100\arcsec~and 315\arcsec~along the slice. This plot shows that the majority of the radio emission is contained within the 8 $\mu$m emission indicating that the ionized gas lies primarily interior to the infrared shell (thermal dust). Figure \ref{fig10} shows the same slice across BD52, but for the 3.6, 4.5, and 5.8 $\mu$m data. The 3.6 and 4.5 $\mu$m bands are more sensitive to stellar emission, and show only faint continuum emission compared to the 5.8 $\mu$m emission. The 5.8 $\mu$m emission closely follows the structure of the 8 $\mu$m emission in Figure \ref{fig9}. The lower intensity emission between the two emission peaks at 5.8 and 8 $\mu$m can be attributed to hot dust. The difference between this emission and the emission peaks at 5.8 and 8 $\mu$m is likely due to the addition of PAH features in the peaks, which delineate the edge of the PDR around the cluster.

\subsubsection{DBCL23}

\begin{figure}
\epsscale{1.0}
\plotone{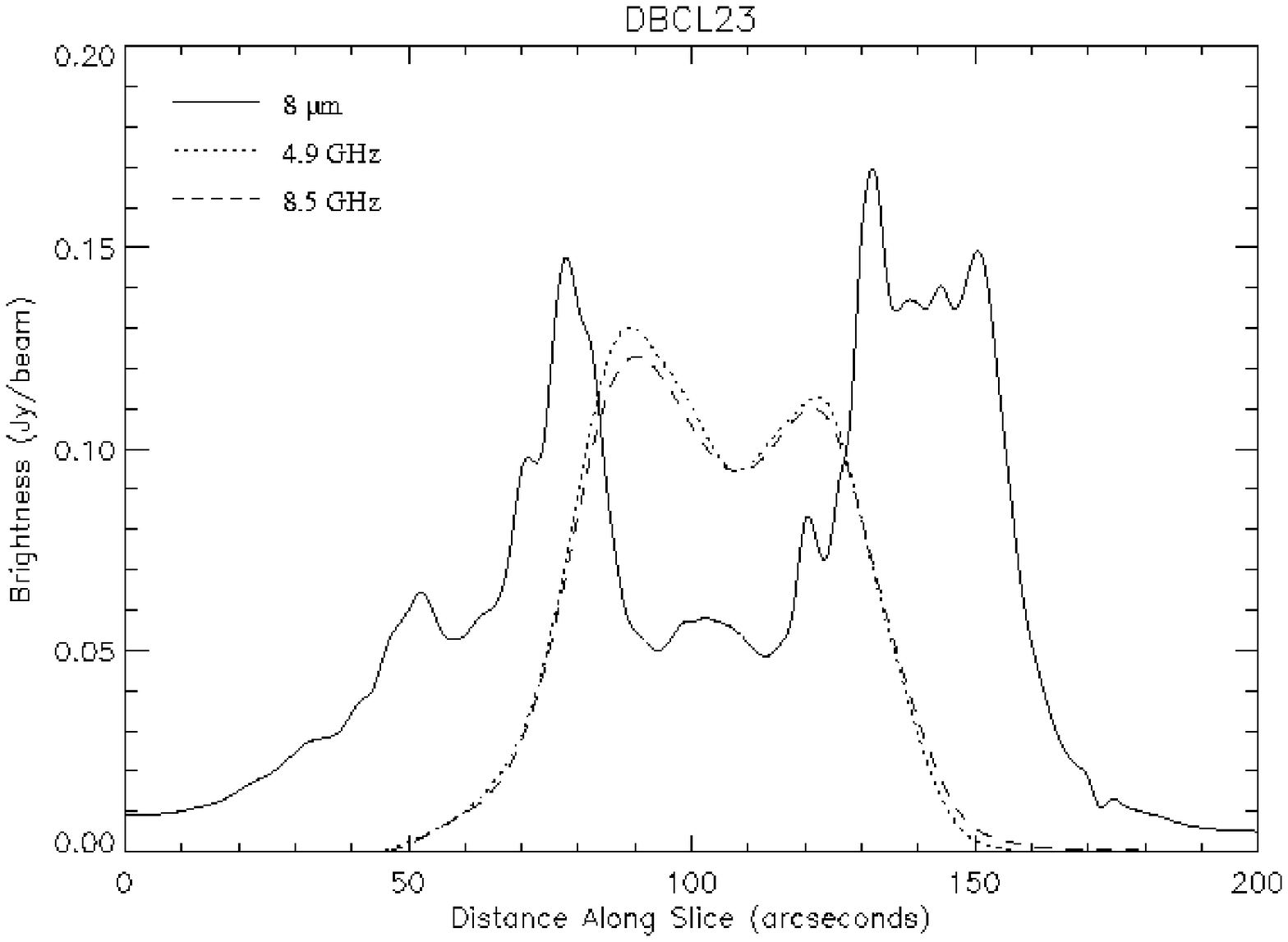}
\caption{DBCL23 slice near DEC = 44\degr15\arcmin15\arcsec~at 8 $\mu$m (solid line), 4.9 GHz (dotted line), and 8.5 GHz (dashed line). The position of the slice is shown exactly in Figure \ref{fig7}. The 8 $\mu$m intensity has been scaled down by a factor of 5$\times$10$^3$. 
\label{fig11}}
\end{figure}

\begin{figure}
\epsscale{1.0}
\plotone{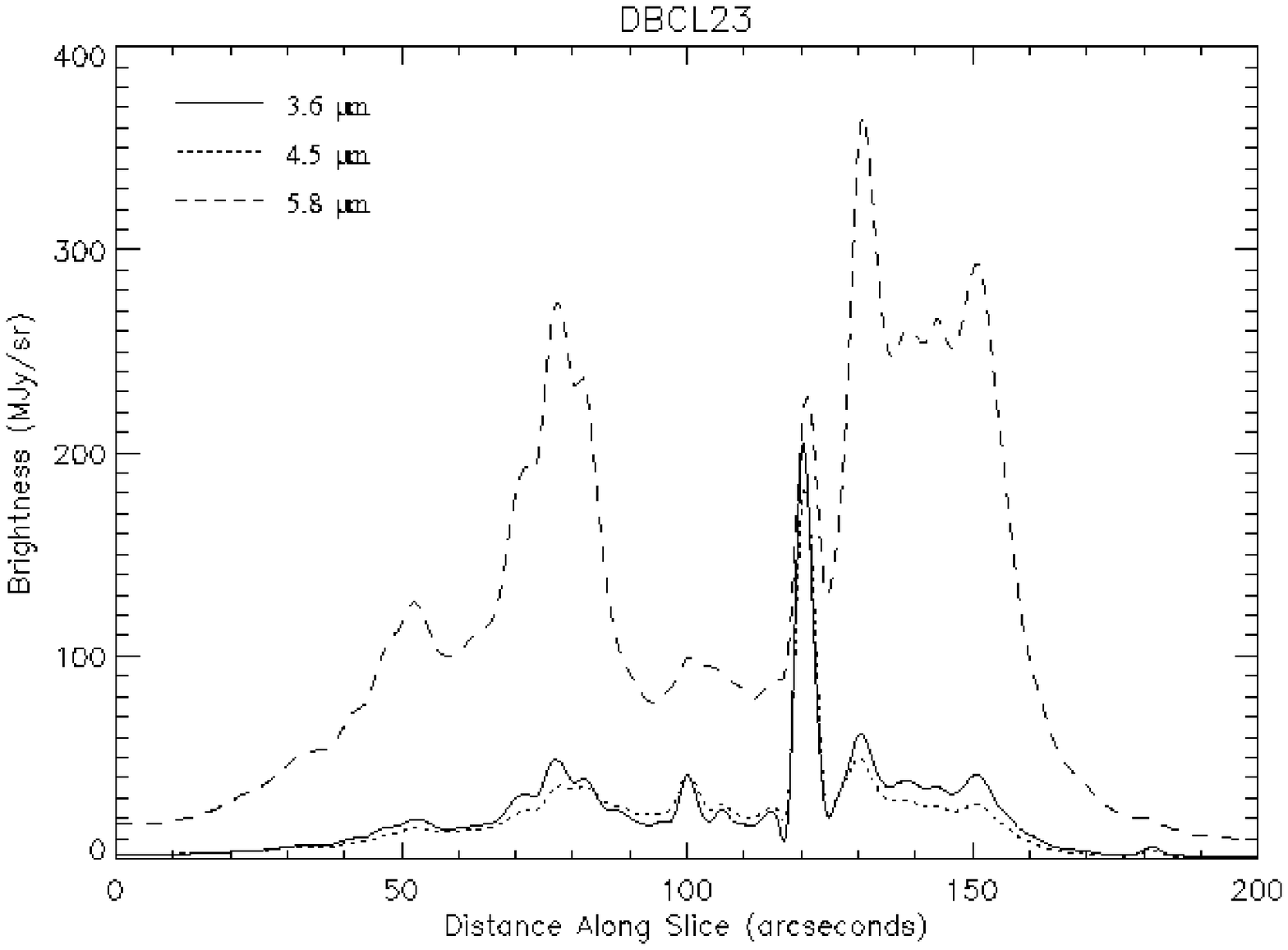}
\caption{DBCL23 slice near DEC = 44\degr15\arcmin15\arcsec~at 3.6 (solid line), 4.5 (dotted line), and 5.8 $\mu$m (dashed line). The position of the slice is shown exactly in Figure \ref{fig7}.
\label{fig12}}
\end{figure}

Figure \ref{fig11} shows a slice through the shell of DBCL23 at 4.9 GHz, 8.5 GHz (convolved to 4.9 GHz) and 8 $\mu$m. The location of this slice is shown in Figure \ref{fig7}. A pair of peaks in the radio emission occur at $\sim$90\arcsec~and 120\arcsec~along the slice. The 8 $\mu$m emission also shows double peaks that occur at $\sim$65\arcsec~and 130\arcsec~along the slice. The majority of the radio emission is contained within the 8 $\mu$m emission as it is in BD52 indicating that the ionized gas lies interior to the thermal dust. Figure \ref{fig12} shows the same slice across DBCL23, but at 3.6, 4.5, and 5.8 $\mu$m. The 3.6 and 4.5 $\mu$m emission roughly follows the 5.8 $\mu$m emission, although the source is significantly fainter at these wavelengths. As in BD52, the 5.8 $\mu$m closely resembles the 8 $\mu$m emission. The dip in intensity between the emission peaks at these wavelengths is likely due to hot dust with the omission of PAHs. The PAHs are traced by the 5.8 and 8 $\mu$m emission in the two emission peaks. The spike at 120\arcsec~seen in all three bands in Figure \ref{fig12} represents stellar emission.

\subsubsection{BD84}

\begin{figure}
\epsscale{1.0}
\plotone{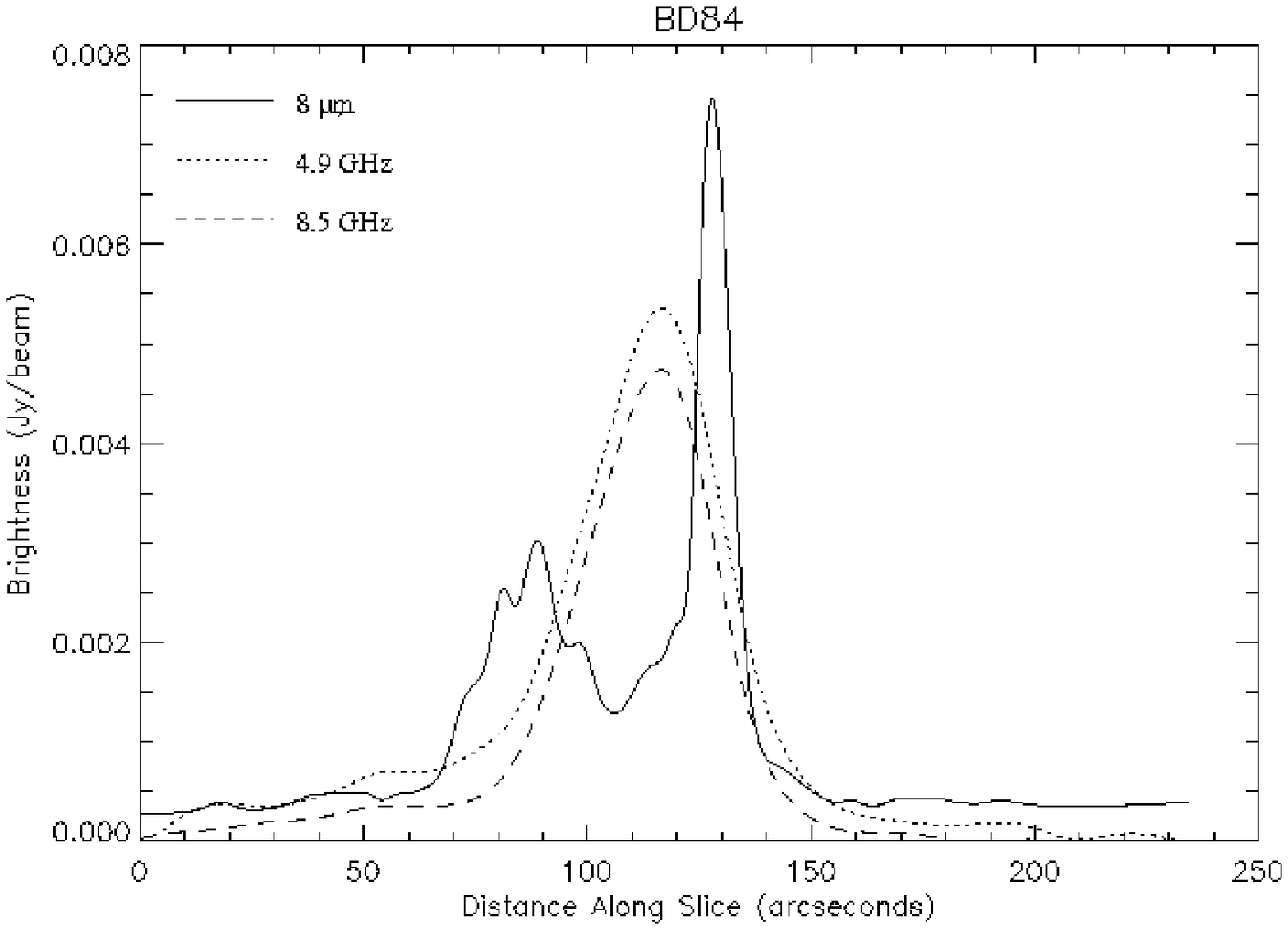}
\caption{BD84 slice at DEC = 0\degr44\arcmin33\arcsec~at 8 $\mu$m (solid line), 4.9 GHz (dotted line), and 8.5 GHz (dashed line). The position of the slice is shown exactly in Figure \ref{fig4}. The 8 $\mu$m intensity has been scaled down by a factor of 2$\times$10$^4$. 
\label{fig13}}
\end{figure}

\begin{figure}
\epsscale{1.0}
\plotone{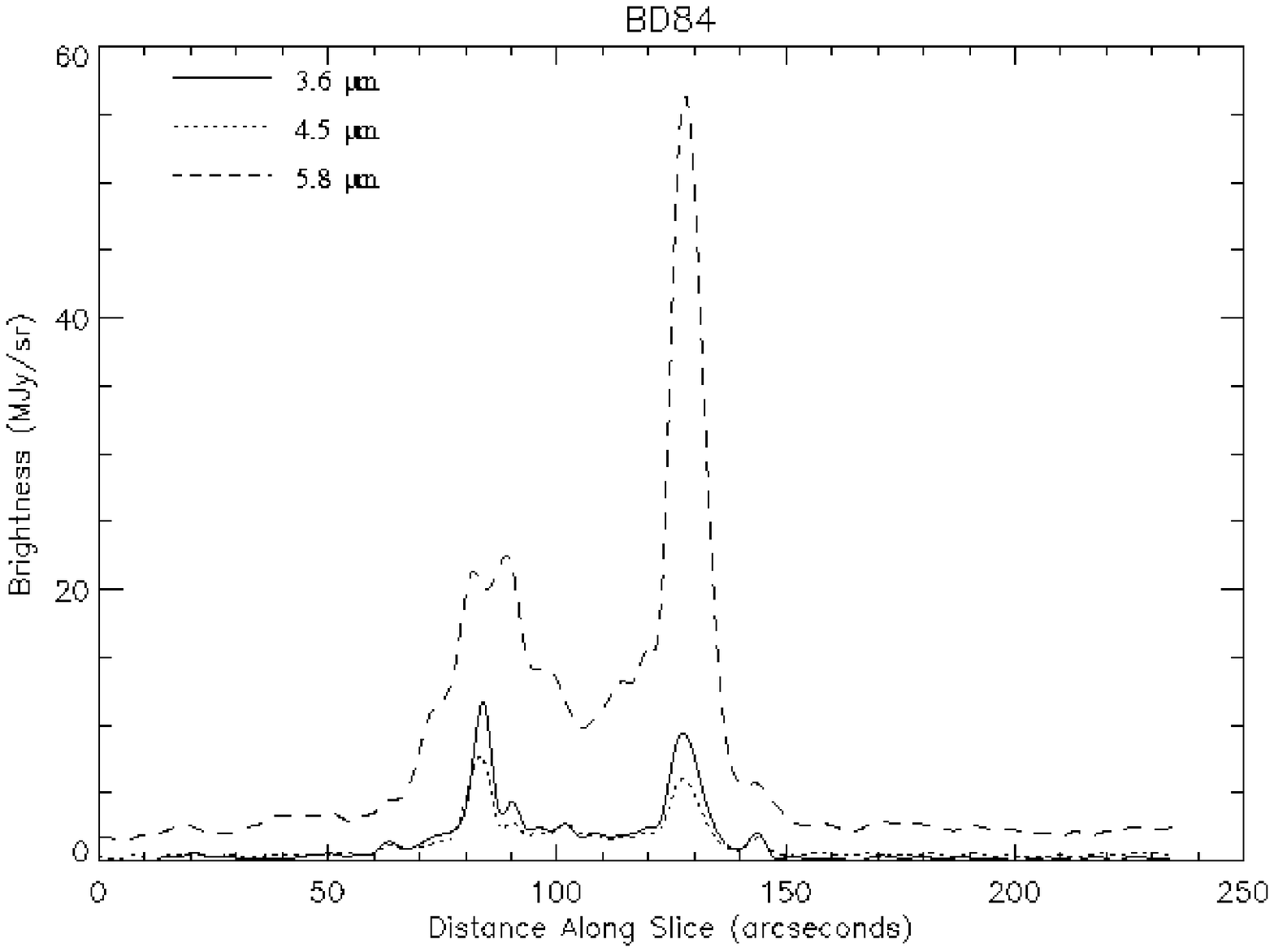}
\caption{BD84 slice at DEC = 0\degr44\arcmin33\arcsec~at 3.6 (solid line), 4.5 (dotted line), and 5.8 $\mu$m (dashed line). The position of the slice is shown exactly in Figure \ref{fig4}.
\label{fig14}}
\end{figure}

Figure \ref{fig13} shows a slice across the source BD84 at 4.9 GHz, 8.5 GHz (convolved to 4.9 GHz) and 8 $\mu$m. The location of this slice is shown in Figure \ref{fig4}. The radio emission for this source has a single peak near $\sim$115-120\arcsec~along the slice, and a FWHM of $\sim$30\arcsec. The 8 $\mu$m emission has peaks on either side of the radio emission, $\sim$90\arcsec~and 130\arcsec. A possible interpretation is that the H II emission indicated by the peak is surrounded by a shell of PAHs and warm dust indicated by the two infrared emission peaks. As is evident in Figure \ref{fig4}, the radio continuum emission is more extended than the infrared emission in BD84. This is only slightly apparent in the slices in Figure \ref{fig13} where the 4.9 GHz emission is visible above the 8 $\mu$m emission just outside the two peaks. The different morphology of BD84 implies there is more ionized gas in this cluster system, and it is not all contained within the PDR. Figure \ref{fig14} shows the same slice across BD84 at 3.6, 4.5, and 5.8 $\mu$m.  It can be seen in Figure \ref{fig14} that the 5.8 $\mu$m band follows the 8 $\mu$m emission very closely. The 3.6 and 4.5 $\mu$m bands exhibit the same general structure, although the second peak at $\sim$130\arcsec~is much smaller, relatively. 

\subsubsection{Slice Comparison}

The double peaked radio emission for the cluster candidates BD52 and DBCL23 implies a shell-like morphology for these H II regions. BD84 differs in this respect with a single peak of radio emission near the cluster surrounded by more diffuse extended emission. Lower intensity emission at 5.8 and 8 $\mu$m between two emission peaks for all three sources indicates PAHs are probably being destroyed by stellar radiation near the clusters. This also provides more evidence that the infrared emission peaks are due to PAH features, and represent the edges of the PDR. Our simplistic analysis for these three sources is consistent with the infrared dust bubble model of \cite{Watson08}.

\subsection{Associated Stellar Sources: Infrared Color-Color Diagrams}

Photometry from {\it Spitzer}/IRAC can be used to identify objects with infrared excess, including evolutionary classes of young stellar objects (YSOs; \citealt{Allen04}). YSOs serve as signposts for sites of ongoing star formation, and can be used to probe age differences in regions of extended star formation. In the simplest case, stars with infrared excess can be identified as candidate YSOs (cYSOs), as the infrared excess is often indicative of a disk structure. Objects which are bright and extended in the 4.5 $\micron$ band, called extended green objects (EGOs), have a high association rate with outflows from massive young stellar objects (MYSOs; \citealt{Cyganowski09}, \citeyear{Cyganowski11}). Sites of massive star formation regions, e.g. in and around massive stellar clusters, can be selected by identifying a population of EGOs.

Using simple assumptions, infrared photometry from {\it Spitzer}/IRAC can be used to construct color-color plots for sources in each field. Figures \ref{fig:bd52midir}-\ref{fig:dbcl23midir} show these plots for all six cluster candidates. In each plot, the left panel shows the $M_{3.6-4.5}, M_{3.6-5.8}$ color-color plot. The right panel shows the distribution of different sources on the 8 $\mu$m image of each cluster candidate. Depending on the location of the sources on the color-color plot, they can be divided into three groups designed to probe different classes of objects: Group I, II and III. Following the procedure outlined in \cite{Zhu09} we construct these groups as follows. Stars with little infrared color difference, e.g. main sequence stars, congregate around $M_{3.6-4.5}, M_{3.6-5.8} = $ 0,0 (i.e., no excess color). Such objects are denoted as Group I in this work. A reddening vector corresponding to A$_{K} = $ 10 is overplotted on each color-color plot. Fore- and background stars affected by differential reddening will follow the reddening vector, and are henceforth denoted as Group III. As stated earlier, excess emission at 5.8 $\micron$ can be indicative of strong PAH emission from circumstellar emission, e.g. a disk around a YSO, and are here referred to as Group II objects. In color-color space, these objects occupy the space defined by $M_{3.6-5.8}$ $>$ 0.9  mag and -0.25 mag $< M_{3.6-4.5} <$ 0.25 mag. Stars with infrared colors in between Groups II and III (see, for example, the left panel of Figure \ref{fig:bd65midir}) are also likely cYSOs.

Active galactic nuclei can contaminate infrared color-color diagrams, though the contamination rate in this sample is likely low given the low latitude of these stellar clusters. Here we discuss the infrared excesses seen toward each of the target sources. Only objects with valid photometry in all four bands are considered. Examination of the distribution of Group II sources indicates that all six clusters are sites of ongoing star formation. Younger, more deeply embedded protostars are undetectable at these wavelengths.

\subsubsection{BD52}
\begin{figure*}
\epsscale{1.0}
\plottwo{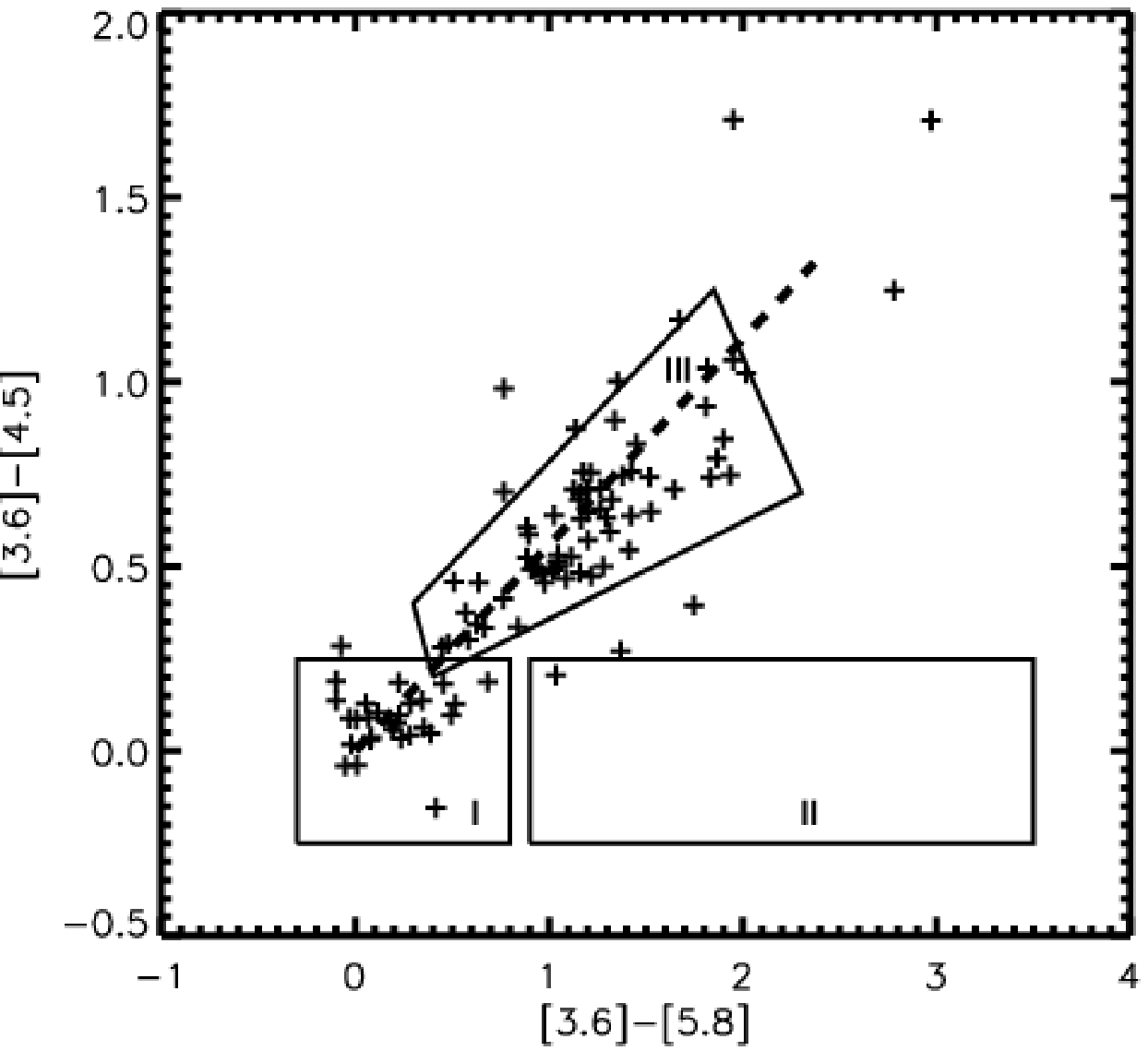}{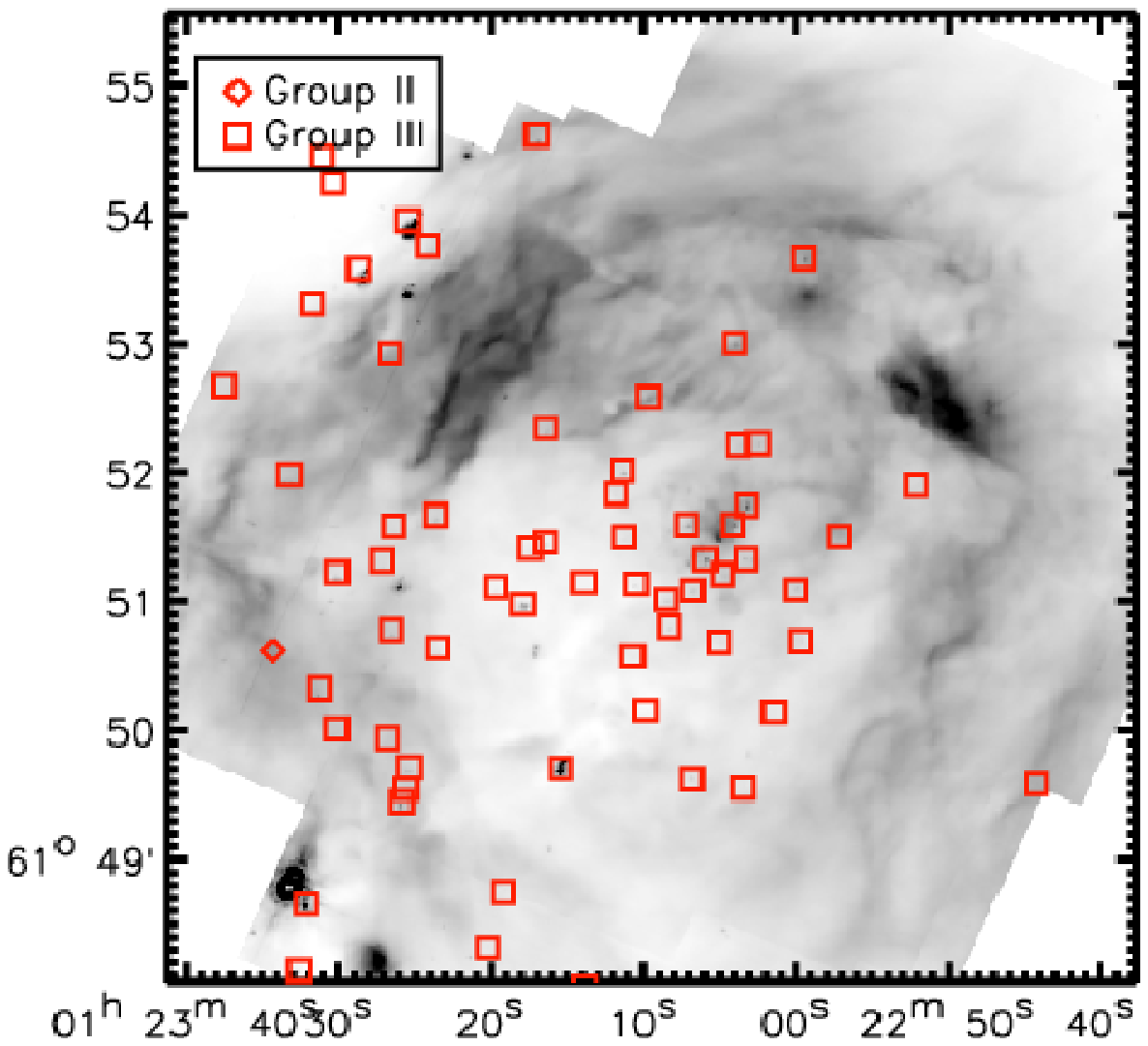}
\caption{Left: IRAC color-color diagram for BD52.  Right: locations of stars showing 5.8 $\mu$m excess (Group II) and strong foreground extinction (Group III), indicated with diamonds and squares, respectively.}\label{fig:bd52midir}
\end{figure*}

\begin{figure*}
\plottwo{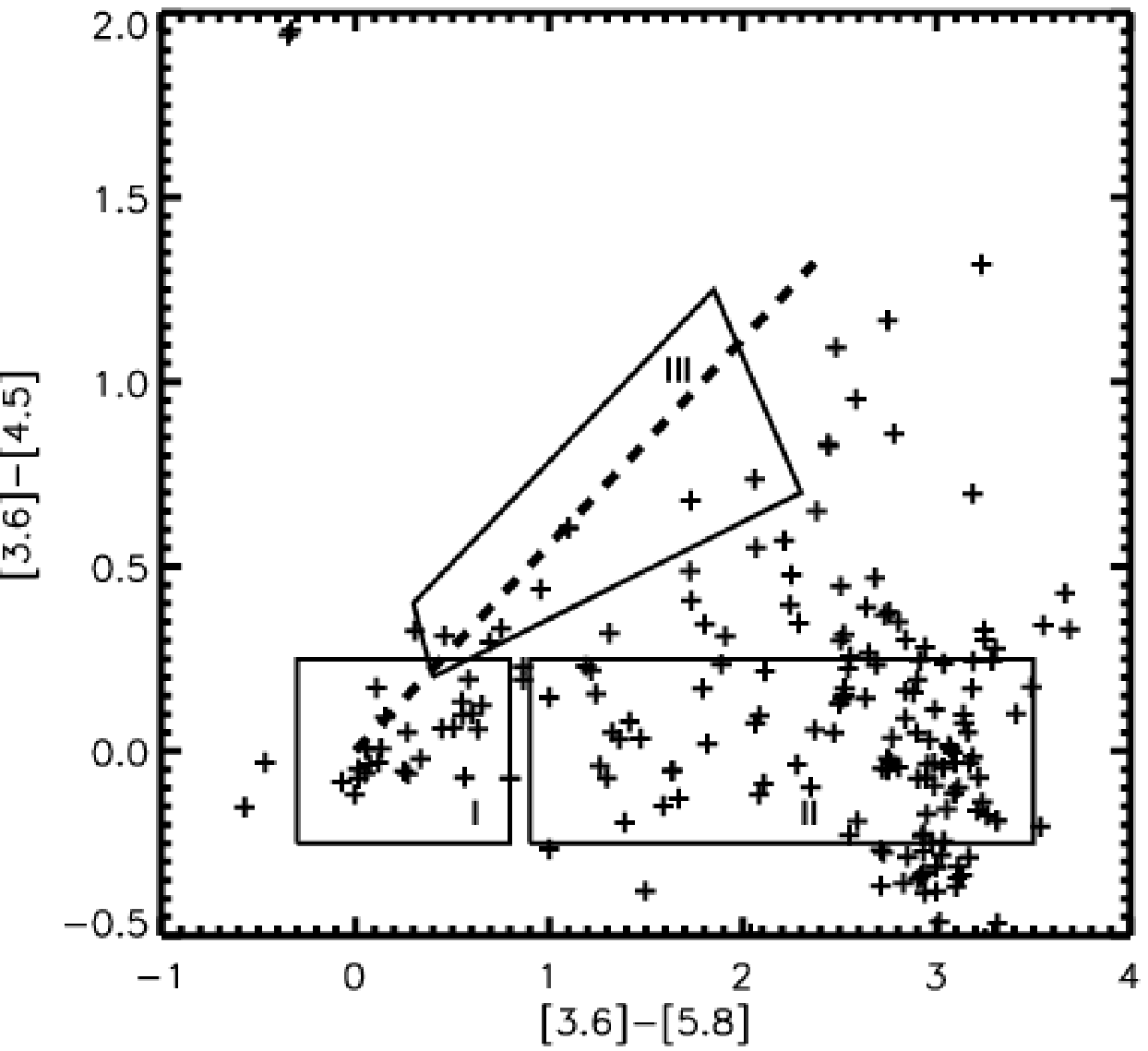}{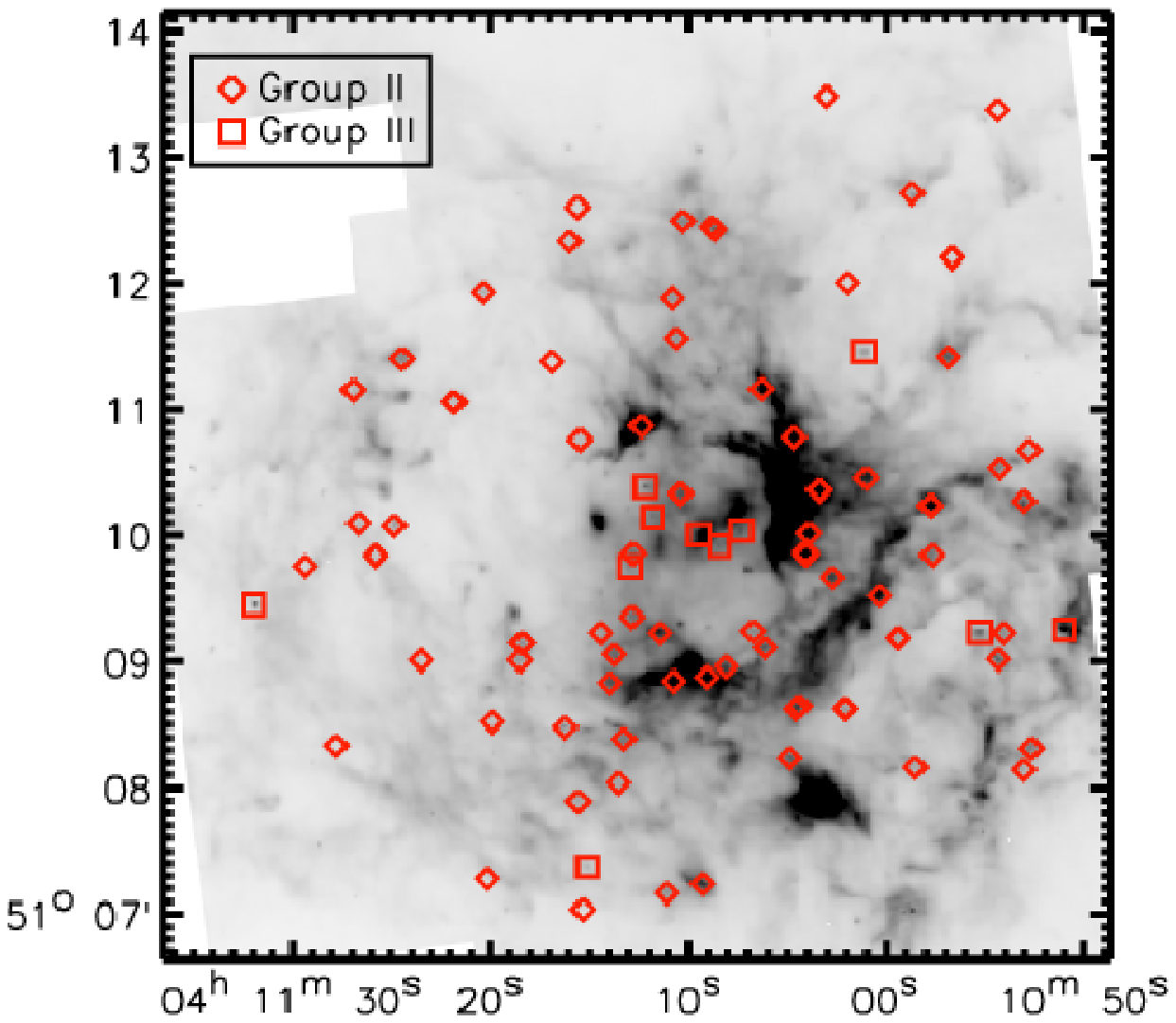}
\caption{Left: IRAC color-color diagram for BD65.  Right: locations of stars showing 5.8 $\mu$m excess (Group II) and strong foreground extinction (Group III), indicated with diamonds and squares, respectively.}\label{fig:bd65midir}
\end{figure*}

\begin{figure*}
\plottwo{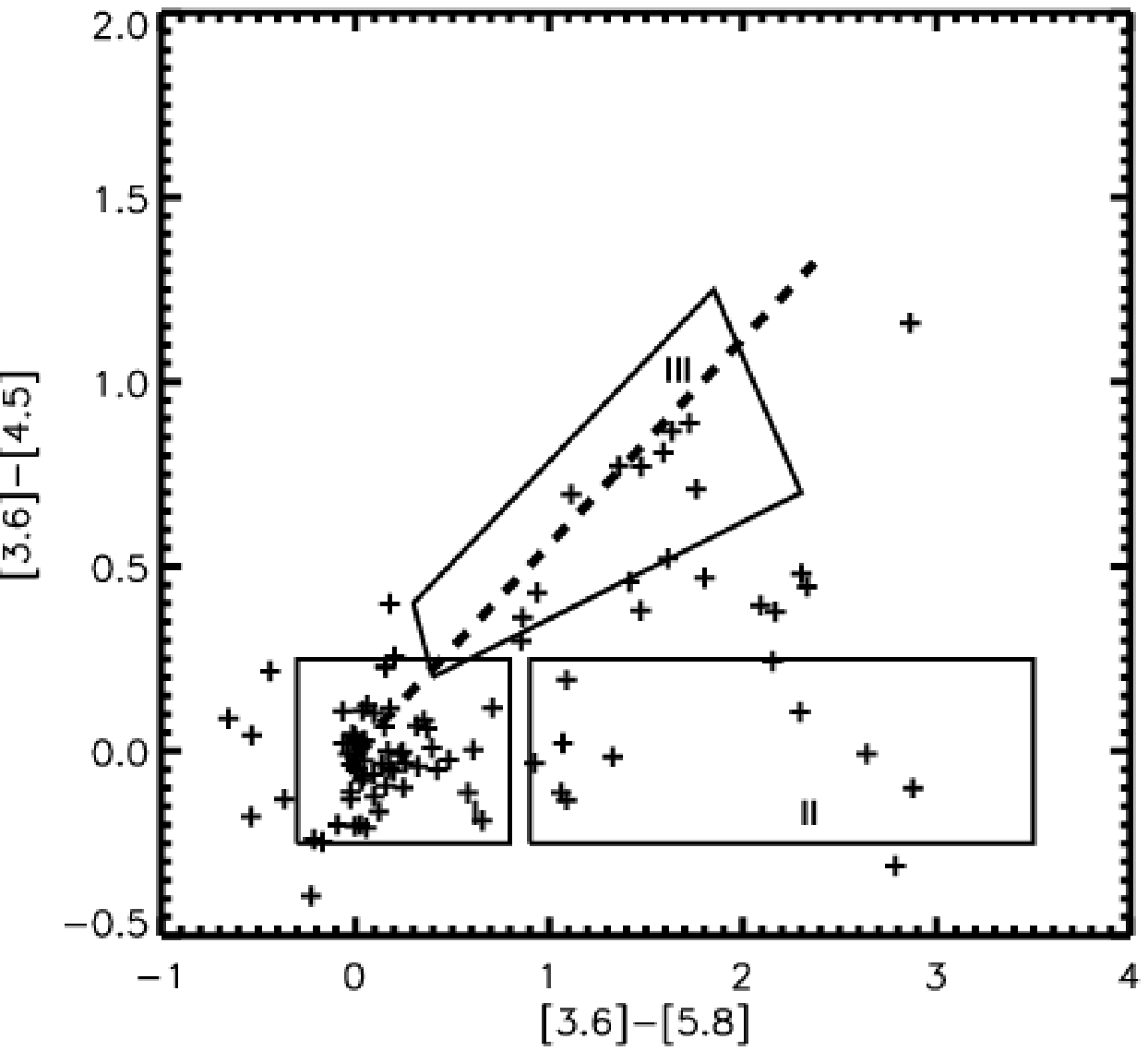}{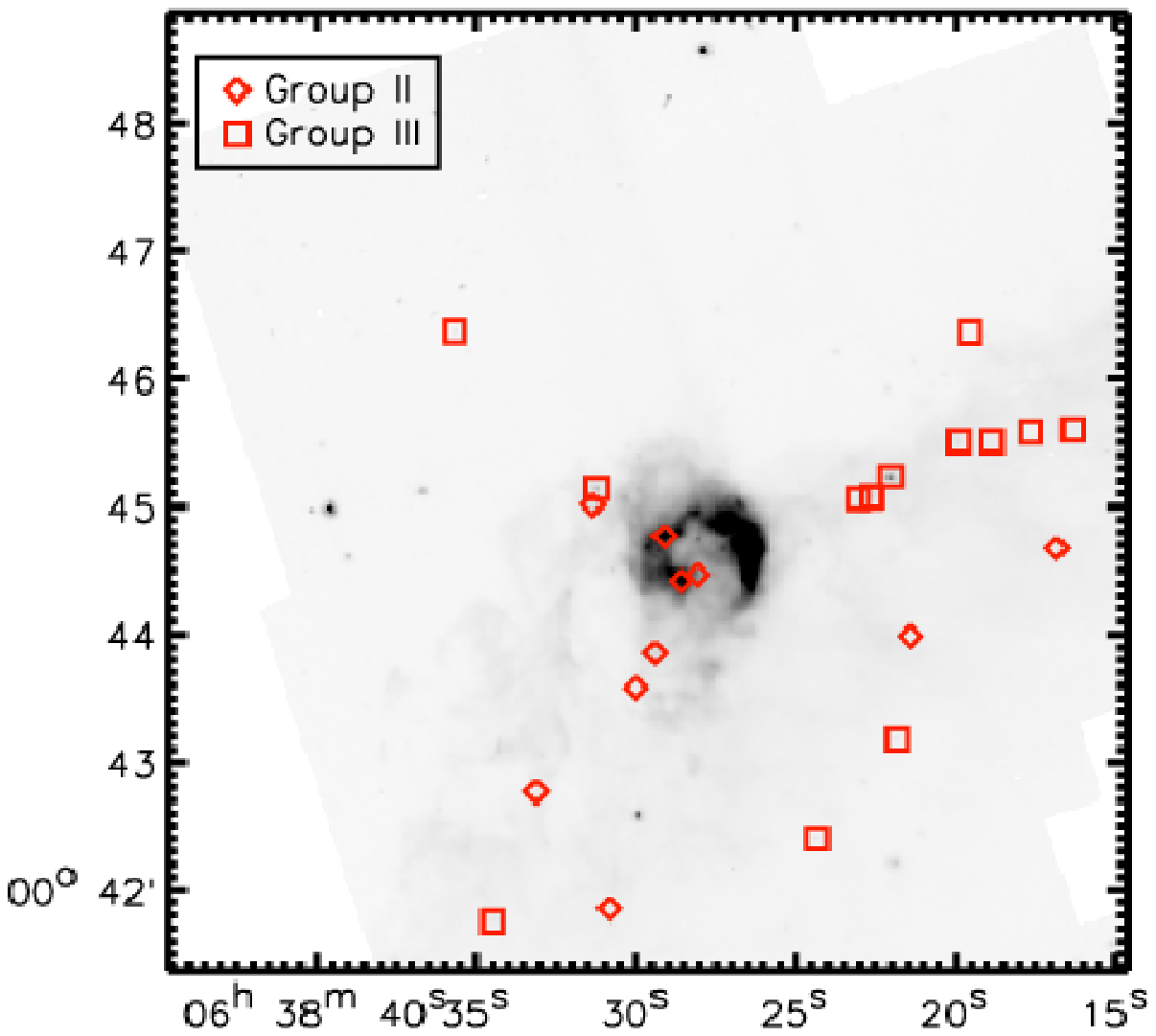}
\caption{Left: IRAC color-color diagram for BD84.  Right: locations of stars showing 5.8 $\mu$m excess (Group II) and strong foreground extinction (Group III), indicated with diamonds and squares, respectively.}\label{fig:bd84midir}
\end{figure*}

\begin{figure*}
\plottwo{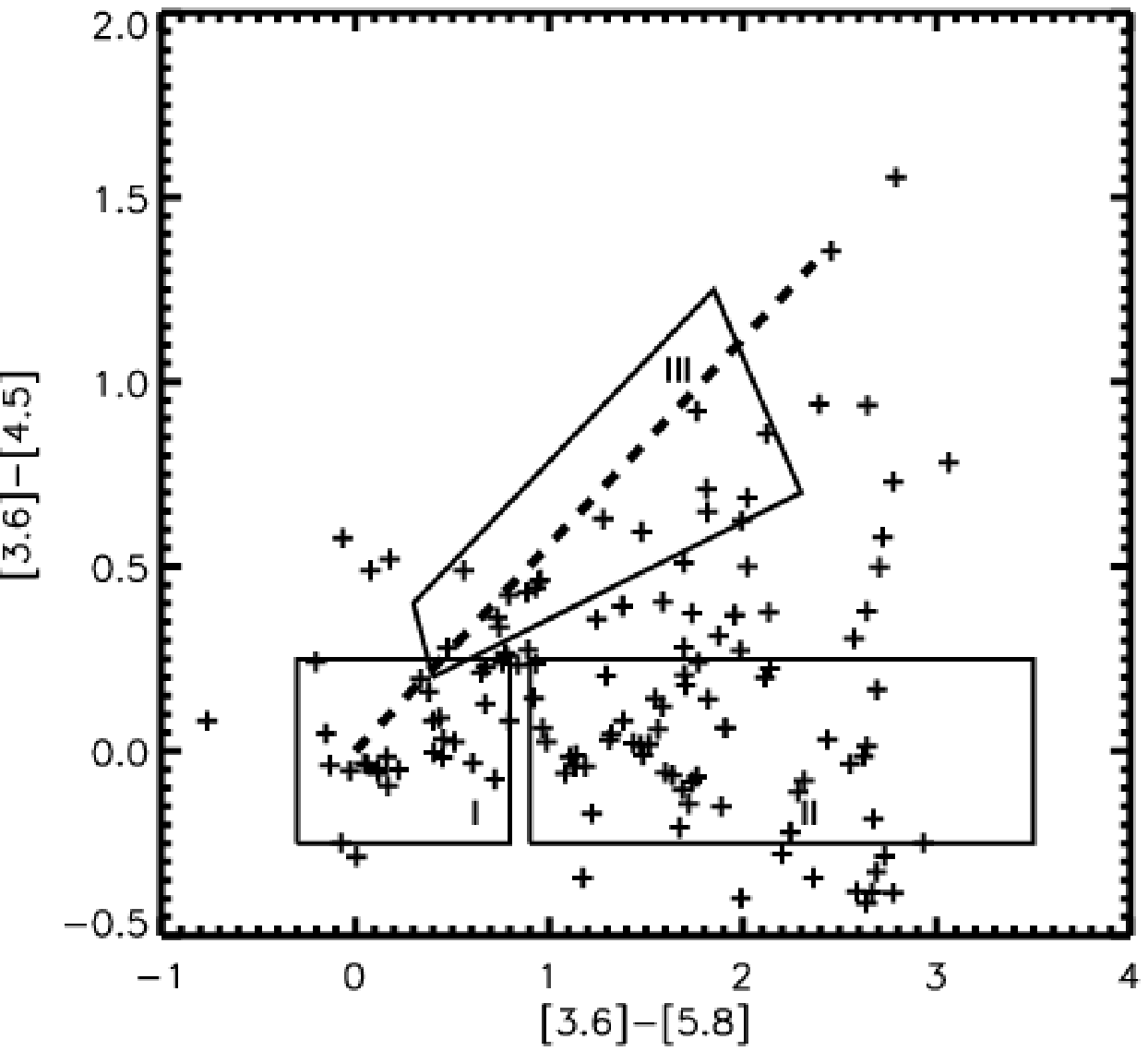}{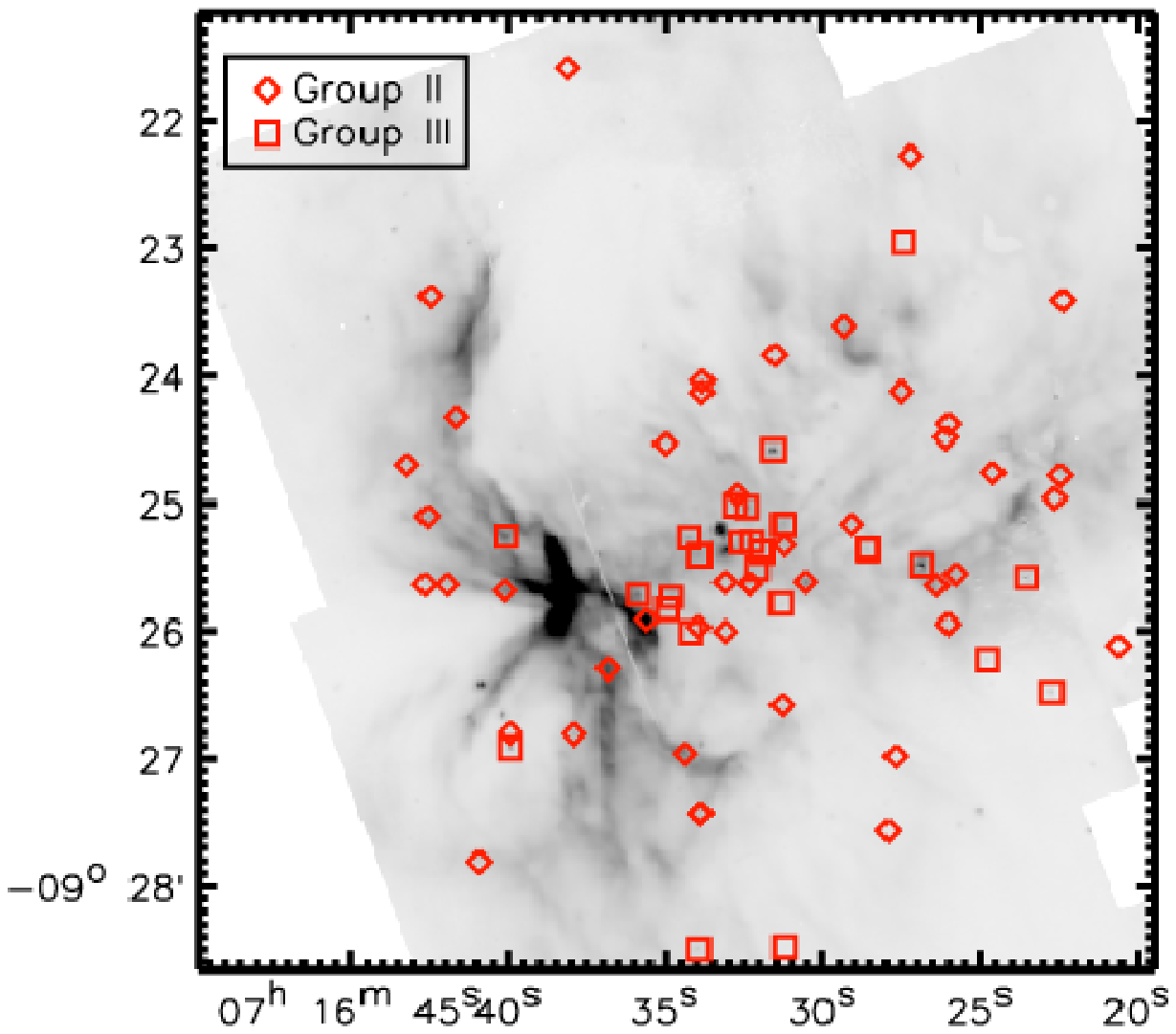}
\caption{Left: IRAC color-color diagram for BD95.  Right: locations of stars showing 5.8 $\mu$m excess (Group II) and strong foreground extinction (Group III), indicated with diamonds and squares, respectively.}\label{fig:bd95midir}
\end{figure*}

\begin{figure*}
\plottwo{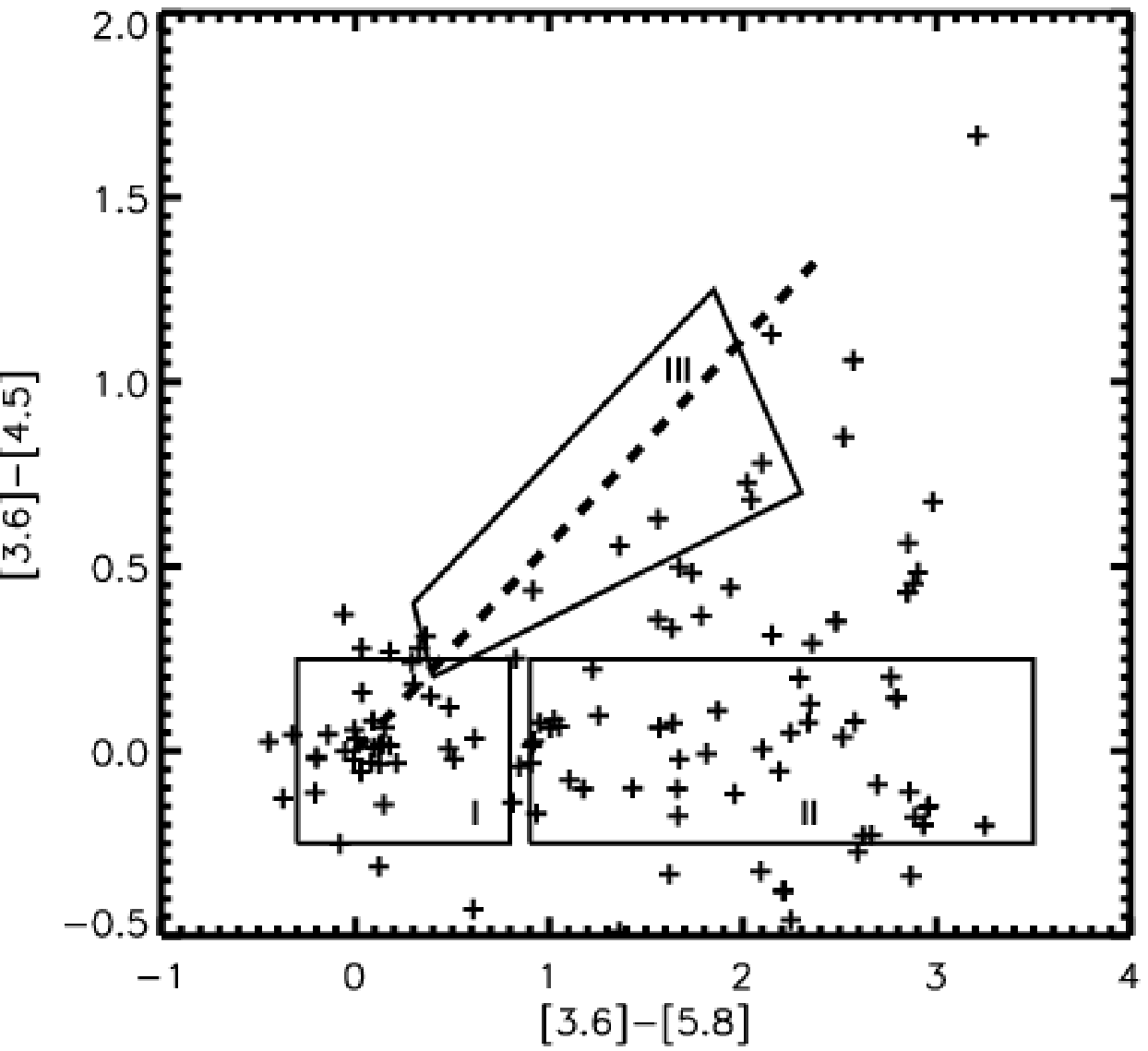}{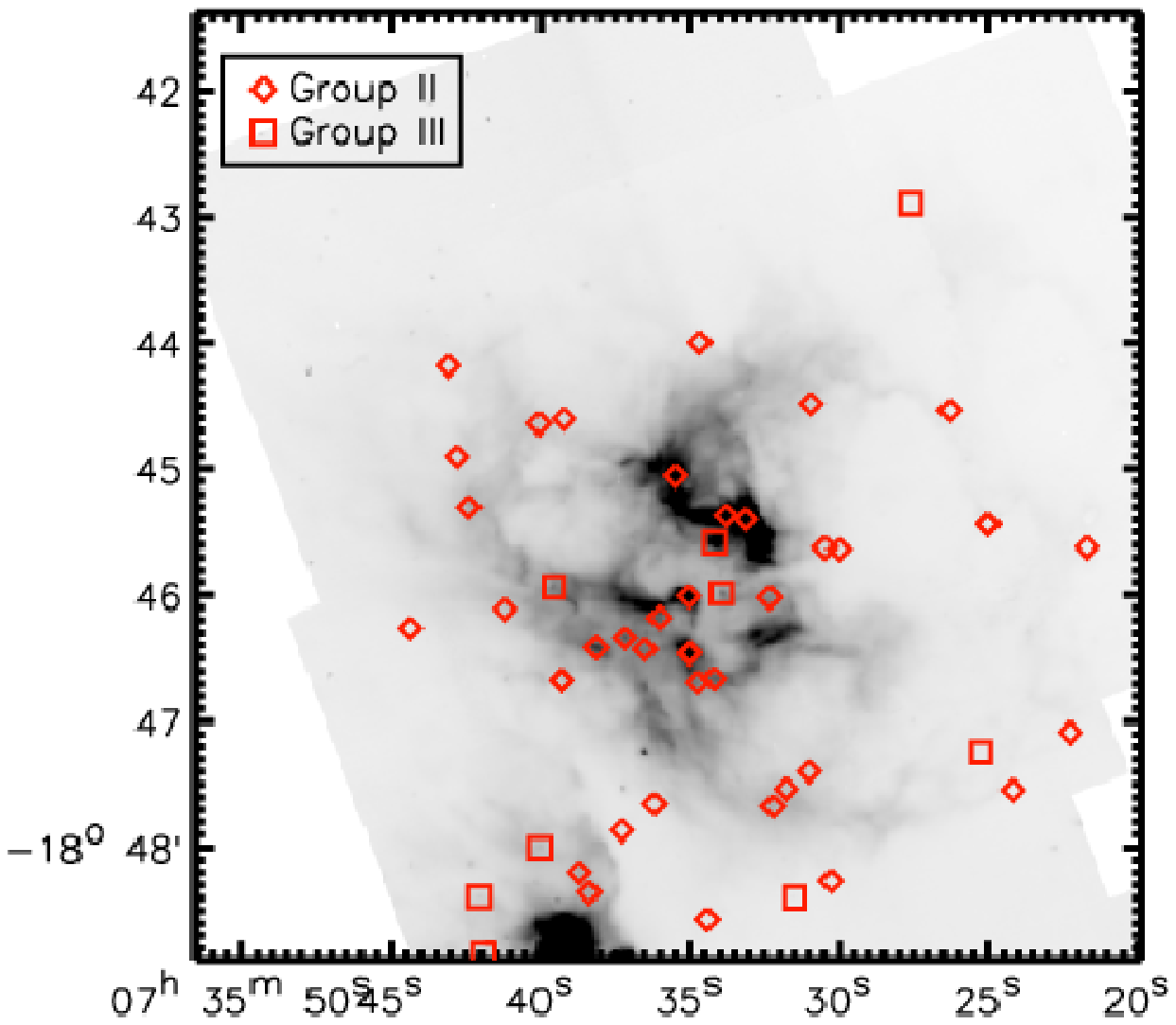}
\caption{Left: IRAC color-color diagram for DB7.  Right: locations of stars showing 5.8 $\mu$m excess (Group II) and strong foreground extinction (Group III), indicated with diamonds and squares, respectively.}\label{fig:db7midir}
\end{figure*}

\begin{figure*}
\plottwo{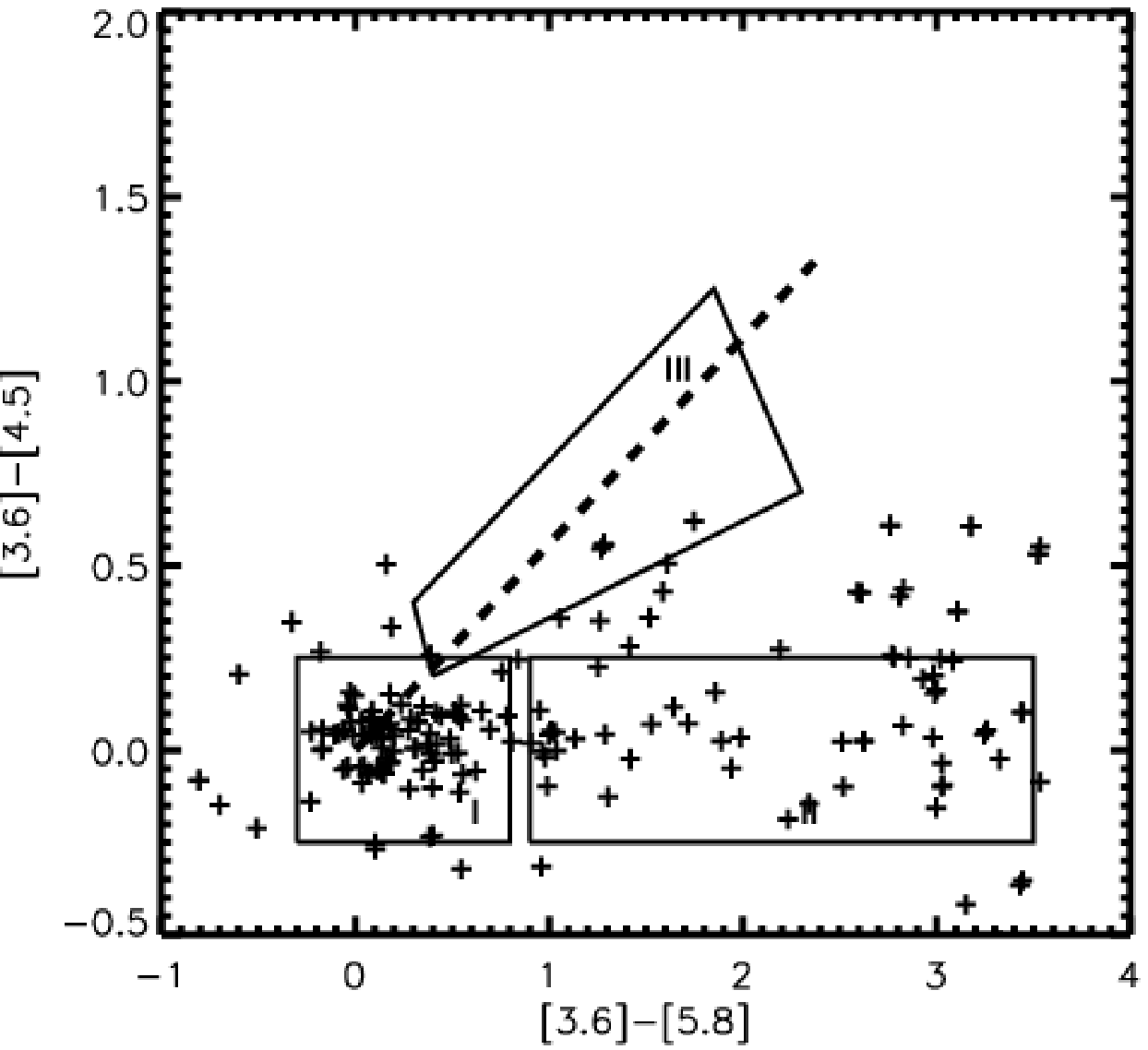}{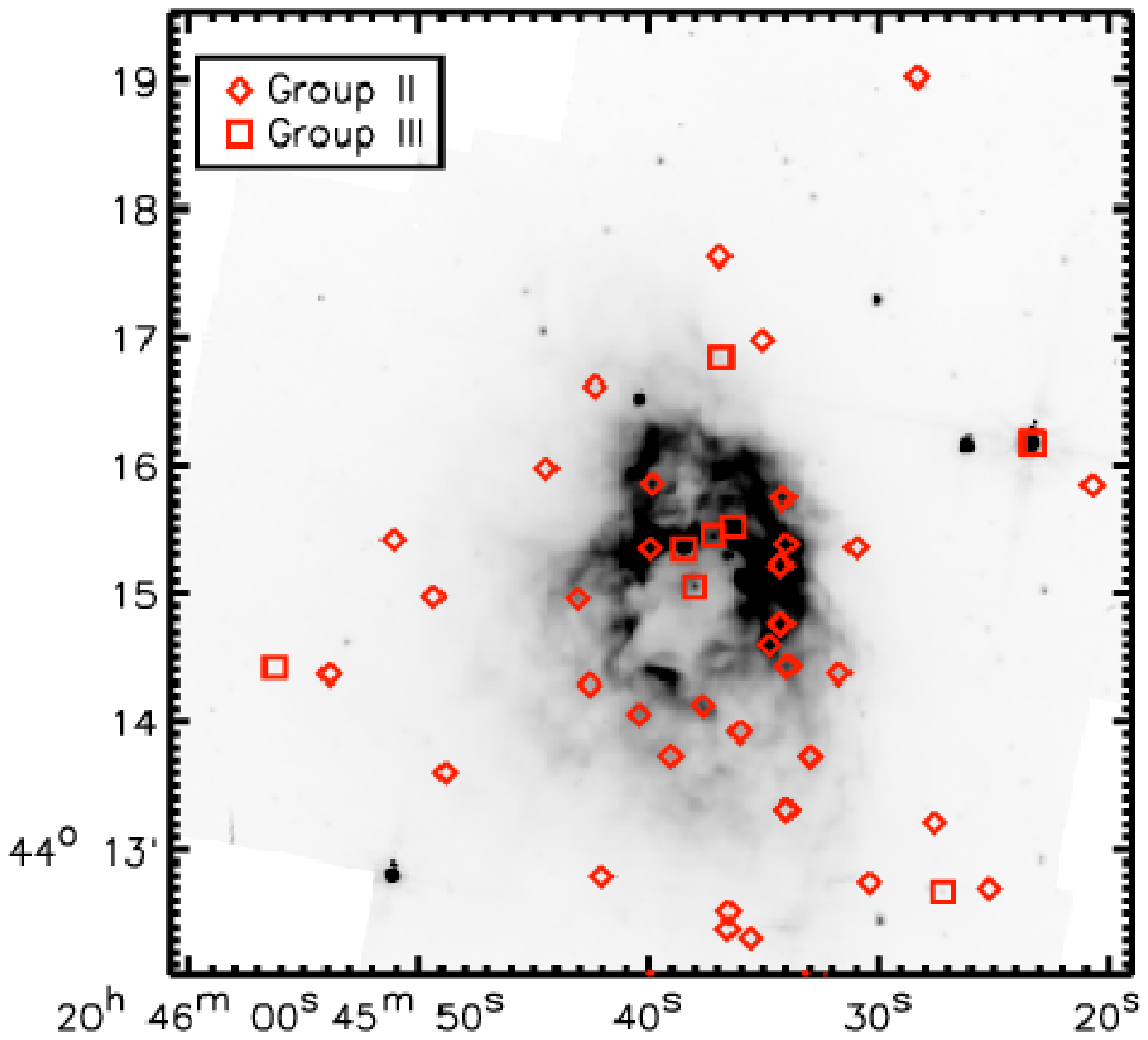}
\caption{Left: IRAC color-color diagram for DBCL23.  Right: locations of stars showing 5.8 $\mu$m excess (Group II) and strong foreground extinction (Group III), indicated with diamonds and squares, respectively.}\label{fig:dbcl23midir}
\end{figure*}

The IRAC color-color diagram of BD52 is shown in the left panel of Figure \ref{fig:bd52midir}. The spatial distribution of Group II and Group III stars is displayed in the right panel. Group III stars are fairly evenly distributed across the face of the cluster, though a dearth of these objects are seen in the direction of bright 8.0 $\micron$ emission peaks. As these sources are distributed along the reddening vector, it is likely that their colors are affected by varying extinction. The 8.0 $\micron$ emission indicates regions of high local extinction, making it difficult to observe background stars in these areas. Only one object appears in the Group II range of colors. This object (shown as a diamond in the right panel of Figure \ref{fig:bd52midir}) is located at the edge of the shell-like 8 $\mu$m emission.

\subsubsection{BD65} 
Figure \ref{fig:bd65midir} displays the IRAC color-color diagram of BD65, left, along with the spatial distribution of objects with infrared excess, right. Group II objects, indicated by a diamond in the right panel of Figure \ref{fig:bd65midir} appear to be distributed preferentially along ridges of the bright, extended 8.0 $\mu$m emission surrounding BD65. The few Group III objects in this region are associated with point sources in the 8.0 $\mu$m image, suggesting that the majority are foreground stars distributed along the line of sight. Stars farther from the reddening vector, falling mostly into Group II, are indistinguishable from cYSOs.

\subsubsection{BD84}
The left panel of Figure \ref{fig:bd84midir} shows the color-color diagram for the sources detected, while the right panel shows the distribution of objects overlaid on the 8 $\mu$m image. The number of infrared sources near BD84 is much lower than for the other target regions. There are only $\sim$35 sources with valid photometry in all four bands; $\sim$100 with valid photometry in 3 bands.  The majority of sources in the field of view are likely to be fore- and background stars. Group III objects follow the reddening vector closely; about half of these sources are located preferentially along a ridge of infrared emission to the West of BD84 and at further distances from the bright infrared emission. Group II sources are more tightly clustered around BD84 and the surrounding bright infrared emission. It is likely that many sources associated with the cluster are undetected. 

\subsubsection{BD95}
The left panel of Figure \ref{fig:bd95midir} shows the color-color diagram for the sources near BD95 and the right panel shows the distribution of the Group II and III sources overlaid on the 8 $\mu$m emission. Both Group II and Group III stars are preferentially located close to the stellar cluster BD95 and along ridges of infrared emission. The majority of these objects are likely cYSOs, highlighting sites of ongoing star formation in and around BD95.

\subsubsection{DB7}
Group II sources in DB7 are located along filaments and knots of bright 8.0 $\micron$ emission, which can be seen in the right panel of Figure \ref{fig:db7midir}. These stars are also located to the SE, in the direction of the stellar cluster DB8. The majority of Group III sources are located fairly far from the reddening vector, and seem to be concentrated in and around the two clusters, making it likely these are cYSOs.

\subsubsection{DBCL23}
Inspection of the right panel of Figure \ref{fig:dbcl23midir} shows that Group II sources are located preferentially towards the bright, diffuse 8.0 $\micron$ emission surrounding DBCL23. The Group III objects located outside of DBCL23 are likely reddened background sources, while those within the cluster are associated with bright point sources, making these sources likely cluster members.

\subsubsection{Overview}
Inspection of the spatial distribution of objects with infrared excess in each of the six clusters presented in this study yield similar results. Each cluster contains candidate YSOs in regions of extended 8.0 $\micron$ emission. It is possible that the distribution of infrared excess sources results from star formation triggered by the expansion of bubble-like H II regions, but it can also be argued that already formed protostellar objects are being exposed by the evolution of the central systems. A more robust analysis of the YSO population of each cluster would be possible at wavelengths between those investigated in this study, e.g. Herschel and/or ALMA observations. These observations would also prove useful as probes of triggered star formation (see \citealt{Zavagno10} for one such study). Maser activity is often associated with shocked emission such as outflows from young stellar objects (e.g. \citealt{Anglada96}); in the case of BD52 the possible EGO is cut off by the edge of the IRAC frame, so a firm association cannot be made.  The maser-YSO association is also pointed out by \cite{Cyganowski08, Cyganowski09} who focus on the association between EGOs and masers. The {\it Spitzer}/IRAC observations of each cluster were inspected by eye for EGOs, yielding no obvious sources. A lack of detection of MYSOs (bright, extended sources in the 4.5 $\mu$m discussed above) is consistent with the general picture for all six clusters: young, partially embedded, intermediate mass stellar clusters. The lack of MYSO detection also implies the candidate clusters are not sites of ongoing massive star formation. For DBCL23, this result is consistent with \cite{Comeron05}'s confirmation that it is not a massive star forming site (\S2.6).

\subsection{Comparison to Known Young, Massive Clusters in the Galaxy}

Table \ref{tbl-5} lists physical properties of 14 young, massive
clusters in the Galaxy that have been well-studied with which the properties of the candidate clusters can be compared. The properties most
relevant to this study are the H II mass and the Lyman continuum
photon flux, because they quantify the impact of these stellar clusters
on the surrounding ISM. It is apparent that in general the average
mass of ionized gas implied by the radio observations for the candidate sources
is below the well-known massive clusters. With the exception of
BD65, all other cluster candidate sources have ionized masses of 7-70
M$_{\sun}$. BD84 and DBCL23 are at the upper end of this range with
ionized masses of 70 and 56 M$_{\sun}$, which are comparable to the
ionized mass associated with Westerlund 1 (\citealt{K-D07}).
BD65 stands out as having a very large ionized mass of 2800 M$_{\sun}$. This mass is
comparable to that of the W49a and Quintuplet H II regions
(\citealt{de Pree97,Lang97}, \citeyear{Lang01a}).

The cluster candidate sources have $\sim$2-3 orders of magnitude smaller Lyman continuum photon fluxes when compared to these other well-studied massive Galactic clusters, which have a typical value $\ge$ 10$^{50}$ s$^{-1}$. Our largest N$_{Lyc}$ value is 3.5$\times$10$^{49}$ s$^{-1}$ for BD65, and even this value is almost an order of magnitude less than the next closest N$_{Lyc}$ value of $\sim$10$^{50}$ s$^{-1}$ for NGC 3603 (\citealt{de Pree99}). The candidate cluster DBCL23 is also on the higher end of our N$_{Lyc}$ range with a value of 2.7$\times$10$^{48}$ s$^{-1}$. This is comparable to DBS2003 45 which had a Lyman continuum flux of 8.2$\times$10$^{48}$ s$^{-1}$ (\citealt{Zhu09}). 

The impact on the ISM from massive clusters such as Arches and Quintuplet results in sizable structures 2 pc and 10 pc in size (\citealt{Lang97},\citeyear{Lang01a}). The size of the H II regions around the cluster sources BD65 and BD84 fall into this range. However, the linear radii of the H II regions around the rest of the candidates are more analogous to the sizes of the actual stellar clusters for some of the comparison clusters (e.g. RSGC 1 \&~2; \citealt{Davies08}). 

Past studies have predicted the Galaxy to harbor about 100 massive ($\sim$10$^{4}$ M$_{\sun}$) stellar clusters similar to the comparison clusters (\citealt{H-P08}). However, using the evidence stated above, we conclude that the candidate clusters do not represent the most massive stellar clusters in the Galaxy, but are rather intermediate mass clusters. \cite{Messineo09} extrapolated the number of known clusters with a mass larger than 10$^{4}$ M$_{\sun}$ to conclude that there are about 200 clusters with masses $>$ 10$^{3}$ M$_{\sun}$. The stellar clusters presented in this study may belong to this group of intermediate mass clusters in the Galaxy. 

\section{CONCLUSIONS}

Observations have been carried out using {\it Spitzer}/IRAC at 3.6, 4.5, 5.8 and 8.0 $\mu$m and VLA radio observations at 8.5 and 4.9 GHz to determine the properties of six candidate stellar clusters and their surrounding H II regions. The following conclusions have been drawn from this study:

\begin{enumerate}
\item Image overlays of all the sources were created to compare the distribution of infrared and radio emission. The 3.6 and 8 $\mu$m emission was found to be more diffuse than the radio emission in the majority of the cluster sources. In addition, the peak intensity in the radio emission is often offset from the peak intensity in the infrared emission. 
\item Radio continuum properties of the H II regions led to an estimate of the sources' ionized mass and Lyman continuum photon flux. BD65 is the most massive of the candidate clusters having an ionized mass of $\sim$2800 M$_\sun$ and N$_{Lyc}$ $\sim$10$^{49}$. DBCL23 was also found to have a fairly large ionizing flux of $\sim$10$^{48}$. The physical parameters of the sources from the radio continuum were compared to H II parameters of other well-studied young, massive Galactic clusters.
\item A more complete multiwavelength analysis of the morphology in BD52, DBCL23, and BD84 was carried out. The emission mechanisms in the clusters and surrounding ISM could be explored by investigating the intensity of emission at different wavelengths across a slice through the sources. A general picture emerged from this analysis: ionized gas surrounded by bright warm dust and PAH emission. This is consistent with other studies of infrared dust bubbles (\citealt{Watson08}).
\item Infrared photometry of stellar members of the candidate clusters was used to search for candidate YSOs. The existence of YSOs provides evidence of ongoing star formation. Candidate YSOs were found in all the cluster candidates using color-color plots. A lack of detection of MYSOs is consistent with the general picture for all six clusters: young, partially embedded, intermediate mass stellar clusters.
\item Comparison with other young, massive clusters (e.g. Arches, Quintuplet, Westerlund 1) leads to the conclusion that the stellar cluster candidates presented in this study are not massive ( $>$ 10$^4$ M$_\sun$) clusters. More likely, they represent examples of Galactic intermediate mass ( $>$ 10$^3$ M$_\sun$) stellar clusters.
\end{enumerate}

\section{ACKNOWLEDGEMENTS}

The authors would like to thank Maria Messineo, Qingfeng Zhu, Ben Davies and Thomas Zimmermann who were involved in the initial stages of this project. CCL would like to acknowledge the University of Iowa Math and Physical Sciences Funding Program for supporting students Emily Richards and Thomas Zimmermann. This material is based upon work supported by the National Science Foundation under Grant No. 0907934.


\clearpage

\clearpage


\begin{deluxetable}{lccc}
\tabletypesize{\scriptsize}
\tablecaption{Candidate Clusters \label{tbl-1}}
\tablewidth{0pt}
\tablehead{
\colhead{Source} & \colhead{RA,DEC} & \colhead{Distance} & \colhead{Alternative} \\
\colhead{Name} & \colhead{(J2000)} & \colhead{(kpc)} & \colhead{Identifiers}}
\startdata
BD52 & 01 23 06.0, 61 51 32.0 & 1.44${\pm}$0.26\tablenotemark{a} & Sh2-187, LBN 126.70-00.80, IRAS 01195+6136 \\

BD65 & 04 11 09.4, 51 10 01.0 & 9.8\tablenotemark{b} & Sh2-209, G151.59-0.23, IRAS 04073+5102 \\

BD84 & 06 38 27.8, 00 44 39.0 & 8.06${\pm}$0.30\tablenotemark{a} &   Sh2-283  \\

BD95 & 07 16 33.0, -09 25 26.0  & 3.24${\pm}$0.56\tablenotemark{a} &  Sh2-294, LBN 1032  \\

DB7 & 07 35 33.7, -18 45 39.0 & 2.65${\pm}$0.4\tablenotemark{c} &  Sh2-307, ESO560-1, SFO49 (DB8)  \\

DBCL23 & 20 45 37.4, 44 15 14.0 & ${\le}$3.5\tablenotemark{d} &   G84.0+0.8, ECX6-38, 3C 423    \\
\enddata
\tablenotetext{a}{\cite{Russeil07}}
\tablenotetext{b}{\cite{Caplan00}}
\tablenotetext{c}{\cite{Messineo07}}
\tablenotetext{d}{\cite{Comeron05}}
\end{deluxetable}

\begin{deluxetable}{lccc}
\tabletypesize{\scriptsize}
\tablecaption{Table for Radio Images \label{tbl-2}}
\tablewidth{0pt}
\tablehead{
\colhead{Source} & \colhead{Frequency} & \colhead{Resolution} & \colhead{rms}\\
\colhead{Name} & \colhead{(GHz)} & \colhead{(arcsec)} & \colhead{(mJy beam$^{-1}$)}}
\startdata
BD52 & 4.9 & 17.93 x 11.61 & 0.313 \\
BD52 & 8.5 & 8.77 x 6.86 & 0.0365 \\
BD65 & 4.9 & 16.60 x 11.41 & 0.552 \\
BD65 & 8.5 & 10.75 x 6.88 & 0.156 \\
BD84 & 4.9 & 29.15 x 11.78 & 0.0675 \\
BD84 & 8.5 & 14.68 x 7.37 & 0.0425 \\
BD95 & 4.9 & 19.24 x 12.76 & 0.281 \\
BD95 & 8.5 & 12.14 x 6.97 & 0.0591 \\
DB7 & 4.9 & 24.27 x 11.77 & 0.194 \\
DB7 & 8.5 & 15.66 x 6.93 & 0.106 \\
DBCL23 & 4.9 & 24.78 x 12.30 & 0.295 \\
DBCL23 & 8.5 & 15.70 x 6.98 & 0.134 \\
\enddata
\end{deluxetable}

\begin{deluxetable}{lcccccc}
\tabletypesize{\scriptsize}
\tablecaption{Radio Continuum Parameters \label{tbl-3}}
\tablewidth{0pt}
\tablehead{
\colhead{Source} & \colhead{Flux Density at 4.9 GHz} & \colhead{Angular Size\tablenotemark{a}} & \colhead{Distance}\\
\colhead{Name} & \colhead{(mJy)} & \colhead{(arcsec)} & \colhead{(kpc)}}
\startdata
BD52 & 503${\pm}$25 & 250 x 250 & 1.44${\pm}$0.26\tablenotemark{b} \\
BD65 & 4100${\pm}$200 & 350 x 200 & 9.8\tablenotemark{c} \\
BD84 & 42${\pm}$3 & 175 x 120 & 8.06${\pm}$0.30\tablenotemark{b} \\
BD95 & 279${\pm}$20 & 220 x 220 & 3.24${\pm}$0.56\tablenotemark{b} \\
DB7 & 511${\pm}$30 & 200 x 175 & 2.65${\pm}$0.4\tablenotemark{d} \\
DBCL23 & 2440${\pm}$100 & 120 x 140 & ${\le}$3.5\tablenotemark{e} \\
\enddata
\tablenotetext{a}{Approximate angular sizes determined using the full width at zero intensity of slices taken across each source delineated by the 4.9 GHz 3$\sigma$ contours.}
\tablenotetext{b}{\cite{Russeil07}}
\tablenotetext{c}{\cite{Caplan00}}
\tablenotetext{d}{\cite{Messineo07}}
\tablenotetext{e}{\cite{Comeron05}}
\end{deluxetable}

\begin{deluxetable}{lcccccc}
\tabletypesize{\scriptsize}
\tablecaption{Physical Quantities of H II Regions Derived from Radio Continuum \label{tbl-4}}
\tablewidth{0pt}
\tablehead{
\colhead{Source} & \colhead {Linear Radius} & \colhead {Electron} & \colhead {Emisson} & \colhead {Ionized} &\colhead  {Lyman} & \colhead{Equivalent}  \\
\colhead {Name} & \colhead {of Sphere} & \colhead {Density} & \colhead {Measure} & \colhead {Mass} & \colhead {Photon Flux} & \colhead{Spectral} \\
\colhead {} & \colhead {(pc)} & \colhead {(cm$^{-3}$)} & \colhead {(pc cm$^{-6}$)} & \colhead {(M$_{\sun}$)} & \colhead {(photons s$^{-1}$)} & \colhead{Classification\tablenotemark{a}}}
\startdata
BD52 & 1.3 & 37 & 2.7$\times$10$^3$ & 7 & 9.3$\times$10$^{46}$ & B0.5V \\
BD65 & 9.1 & 37 & 2.6$\times$10$^{4}$ & 2800 & 3.5$\times$10$^{49}$ & O3V \\
BD84 & 4.1 & 10 & 8.7$\times$10$^{2}$ & 70 & 2.4$\times$10$^{47}$  & B0V \\
BD95 & 2.5 & 22 & 2.5$\times$10$^{3}$ & 35 & 2.6$\times$10$^{47}$ & B0V \\
DB7 & 1.7 & 43 & 6.4$\times$10$^{3}$ & 22 & 3.2$\times$10$^{47}$ & B0V \\
DBCL23 & 1.6 & 140 & 6.4$\times$10$^{4}$ & 56 & 2.7$\times$10$^{48}$ & O8V \\
\enddata
\tablecomments{Quantities calculated assuming T$_{e}$ = 10,000 K and Y$^+$ = 0.1.}
\tablenotetext{a}{Spectral type estimates of exciting stars are derived from N$_{Lyc}$ using model stellar atmosphere results from \cite{Smith02}} 
\end{deluxetable}

\begin{deluxetable}{lccccccccc}
\tabletypesize{\scriptsize}
\rotate
\tablecaption{Cluster Comparison Parameters \label{tbl-5}}
\tablewidth{0pt}
\tablehead{
\colhead{Source} & \colhead{Distance} & \multicolumn{2}{c}{Galactic Coordinates} & \colhead{Age} & \colhead{Cluster Radius} & \colhead{Cluster Mass} & \colhead{H II Mass} & \colhead{N$_{Lyc}$} & \colhead{References} \\
\colhead{Name} & \colhead{(kpc)} & \colhead{$\ell$} & \colhead{b} & \colhead{(Myr)} & \colhead{(pc)} & \colhead{(1000$\times$M$_{\sun}$)} & \colhead{(M$_{\sun}$)} & \colhead{(s$^{-1}$)} & \colhead{}}
\startdata
Westerlund 1 & 3.55 & 339.55 & -0.4 & 3-5 & 1.0 & 45 & 47-53 & ... & 1,2 \\
Westerlund 2 & 2.8 & 284.27 & -0.34 & 1-3 & ... & 7 & ... & 6$\times$10$^{50}$ & 3,4 \\
RCW49 & 8 & 284.31 & -0.33 & ... & ... & ... & ... & 11.5$\times$10$^{50}$ & 3 \\
RSGC1 & 6.6 & 25.27 & -0.16 & 10-14 & 1.5 & 30 & ... & ... & 5 \\
RSGC2 & 5.8 & 26.19 & -0.07 & 14-20 & 3.2 & 40 & ... & ... & 5 \\
W49a & 11.4 & 43.17 & 0.0018 & 0.3-2 & ... & 50-70 & ${\ge}$ 238 & 2.4$\times$10$^{50}$ & 6,7 \\
Central & 8 & 0.00 & 0.01 & 3-7 & 0.23 & 20 & ... & 3.2$\times$10$^{50}$ & 8,9 \\
Arches & 8 & 0.12 & 0.02 & 2-3 & 0.23 & ${\le}$ 70 & 5480 & 4$\times$10$^{51}$ & 10,11 \\
Quintuplet & 8 & 0.16 & -0.06 & 3-5 & 1.0 & 20 & 240 & 7.9$\times$10$^{50}$ & 9,12 \\
NGC 3603 & 7.6 & 291.63 & -0.53 & 2.5 & 0.7-1.5 & 10-16 & 10000 & 1$\times$10$^{50}$ & 13,14 \\
CL 1806-20 & 8.7 & 10.00 & -0.24 & 3-4.5 & ... & 3 & ... & ... & 15,16 \\
Glimpse 30 & 7.2 & 298.76 & -0.41 & 4-5 & 1.36 & 3 & ... & ... & 8 \\
DBS2003 179 & 7.9 & 347.58 & 0.19 & 2-5 & 0.2 & 7 & ... & ... & 17 \\
DBS2003 45 & 4.5 & 283.88 & -0.91 & 5-8 & ... & 1 & ... & 8.2$\times$10$^{48}$ & 18 \\
\enddata
\tablerefs{(1) {\cite{Brandner08}}; (2) {\cite{K-D07}}; (3) {\cite{Rauw07}}; (4) {\cite{Ascenso07}}; (5) {\cite{Davies08}}; (6) {\cite{H-A05}}; (7) {\cite{de Pree97}}; (8) {\cite{Kurtev07}}; (9) {\cite{Figer99a}}; (10) {\cite{Figer02}}; (11) {\cite{Lang01a}}; (12) {\cite{Lang97}} (13) {\cite{Melena08}}; (14) {\cite{de Pree99}}; (15) {\cite{Bibby08}}; (16) {\cite{Figer05}}; (17) {\cite{Borissova08}}; (18) {\cite{Zhu09}}}


\end{deluxetable}


\begin{thebibliography}{}

\bibitem[Allen et al.(2004)]{Allen04} Allen, L.~E., et al.\ 2004, \apjs, 154, 363
\bibitem[Anglada et al.(1996)]{Anglada96} Anglada, G., Estalella, R., Pastor, J., Rodriguez, L.~F., \& Haschick, A.~D.\ 1996, \apj, 463, 205 
\bibitem[Ascenso et al.(2007)]{Ascenso07} Ascenso, J., Alves, J., Beletsky, Y., \& Lago, M.~T.~V.~T.\ 2007, \aap, 466, 137 
\bibitem[Balser et al.(1995)]{Balser95} Balser, D.~S., Bania, T.~M., Rood, R.~T., \& Wilson, T.~L.\ 1995, \apjs, 100, 371 
\bibitem[Bibby et al.(2008)]{Bibby08} Bibby, J.~L., Crowther, P.~A., Furness, J.~P., \& Clark, J.~S.\ 2008, \mnras, 386, L23
\bibitem[Bica et al.(2003)]{Bica03} Bica, E., Dutra, C.~M., Soares, J., \& Barbuy, B.\ 2003, \aap, 404, 223 
\bibitem[Borissova et al.(2008)]{Borissova08} Borissova, J., Ivanov, V.~D., Hanson, M.~M., Georgiev, L., Minniti, D., Kurtev, R., \& Geisler, D.\ 2008, \aap, 488, 151 
\bibitem[Brandner et al.(2008)]{Brandner08} Brandner, W., Clark, J.~S., Stolte, A., Waters, R., Negueruela, I., \& Goodwin, S.~P.\ 2008, \aap, 478, 137 
\bibitem[Caplan et al.(2000)]{Caplan00} Caplan, J., Deharveng, L., Pe{\~n}a, M., Costero, R., \& Blondel, C.\ 2000, \mnras, 311, 317 
\bibitem[Comer{\'o}n et al.(2005)]{Comeron05} Comer{\'o}n, F., Pasquali, A., \& Torra, J.\ 2005, \aap, 440, 163 
\bibitem[Condon et al.(1998)]{Condon98} Condon, J.~J., Cotton, W.~D., Greisen, E.~W., Yin, Q.~F., Perley, R.~A., Taylor, G.~B., \& Broderick, J.~J.\ 1998, \aj, 115, 1693
\bibitem[Cotera et al.(1996)]{Cotera96} Cotera, A.~S., Erickson, E.~F., Colgan, S.~W.~J., Simpson, J.~P., Allen, D.~A., \& Burton, M.~G.\ 1996, \apj, 461, 750 
\bibitem[Cyganowski et al.(2008)]{Cyganowski08} Cyganowski, C.~J., Whitney, B.~A., Holden, E., et al.\ 2008, \aj, 136, 2391 
\bibitem[Cyganowski et al.(2009)]{Cyganowski09} Cyganowski, C.~J., Brogan, C.~L., Hunter, T.~R., \& Churchwell, E.\ 2009, \apj, 702, 1615 
\bibitem[Cyganowski et al.(2011)]{Cyganowski11} Cyganowski, C.~J., Brogan, C.~L., Hunter, T.~R., Churchwell, E., \& Zhang, Q.\ 2011, \apj, 729, 124 
\bibitem[Davies et al.(2008)]{Davies08} Davies, B., Figer, D.~F., Law, C.~J., Kudritzki, R.-P., Najarro, F., Herrero, A., \& MacKenty, J.~W.\ 2008, \apj, 676, 1016 
\bibitem[de Pree et al.(1997)]{de Pree97} de Pree, C.~G.,Mehringer, D.~M., \& Goss, W.~M.\ 1997, \apj, 482, 307
\bibitem[de Pree et al.(1999)]{de Pree99} de Pree, C.~G., Nysewander, M.~C., \& Goss, W.~M.\ 1999, \aj, 117, 2902 
\bibitem[Dutra \& Bica(2001)]{D-B01} Dutra, C.~M., \& Bica, E.\ 2001, \aap, 376, 434
\bibitem[Dutra et al.(2003)]{Dutra03} Dutra, C.~M., Bica, E., Soares, J., \& Barbuy, B.\ 2003, \aap, 400, 533
\bibitem[Figer et al.(1999a)]{Figer99a} Figer, D.~F., McLean, I.~S., \& Morris, M.\ 1999, \apj, 514, 202 
\bibitem[Figer et al.(1999b)]{Figer99b} Figer, D.~F., Kim, S.~S., Morris, M., Serabyn, E., Rich, R.~M., \& McLean, I.~S.\ 1999, \apj, 525, 750
\bibitem[Figer et al.(2002)]{Figer02} Figer, D.~F., et al.\ 2002, \apj, 581, 258 
\bibitem[Figer et al.(2005)]{Figer05} Figer, D.~F., Najarro, F., Geballe, T.~R., Blum, R.~D., \& Kudritzki, R.~P.\ 2005, \apjl, 622, L49 
\bibitem[Hanson \& Popescu(2008)]{H-P08} Hanson, M.~M., \& Popescu, B.\ 2008, IAU Symposium, 250, 307 
\bibitem[Homeier \& Alves(2005)]{H-A05} Homeier, N.~L., \& Alves, J.\ 2005, \aap, 430, 481
\bibitem[Joncas et al.(1992)]{Joncas92} Joncas, G., Durand, D., \& Roger, R.~S.\ 1992, \apj, 387, 591
\bibitem[Klein et al.(2005)]{Klein05} Klein, R., Posselt, B., Schreyer, K., Forbrich, J., \& Henning, T.\ 2005, \apjs, 161, 361 
\bibitem[Kothes \& Dougherty(2007)]{K-D07} Kothes, R., \& Dougherty, S.~M.\ 2007, \aap, 468, 993 
\bibitem[Kurtev et al. (2007)]{Kurtev07} Kurtev, R., Borissova, J., Georgiev, L., Ortolani, S., \& Ivanov, V.~D.\ 2007, \aap, 475, 209
\bibitem[Lang et al.(1997)]{Lang97} Lang, C.~C., Goss, W.~M., \& Wood, D.~O.~S.\ 1997, \apj, 474, 275 
\bibitem[Lang et al.(2001a)]{Lang01a} Lang, C.~C., Goss, W.~M., \& Morris, M.\ 2001, \aj, 121, 2681
\bibitem[Lang et al.(2001b)]{Lang01b} Lang, C.~C., Goss, W.~M., \& Rodr{\'{\i}}guez, L.~F.\ 2001, \apjl, 551, L143 
\bibitem[Lang et al.(2002)]{Lang02} Lang, C.~C., Goss, W.~M., \& Morris, M.\ 2002, \aj, 124, 2677 
\bibitem[Melena et al.(2008)]{Melena08} Melena, N.~W., Massey, P., Morrell, N.~I., \& Zangari, A.~M.\ 2008, \aj, 135, 878 
\bibitem[Mercer et al.(2005)]{Mercer05} Mercer, E.~P., et al.\ 2005, \apj, 635, 560 
\bibitem[Messineo et al.(2007)]{Messineo07} Messineo, M., Petr-Gotzens, M.~G., Schuller, F., Menten, K.~M., Habing, H.~J., Kissler-Patig, M., Modigliani, A., \& Reunanen, J.\ 2007, \aap, 472, 471 
\bibitem[Messineo et al.(2009)]{Messineo09} Messineo, M., Davies, B., Ivanov, V.~D., Figer, D.~F., Schuller, F., Habing, H.~J., Menten, K.~M., \& Petr-Gotzens, M.~G.\ 2009, \apj, 697, 701
\bibitem[Mezger \& Henderson(1967)]{M-H67} Mezger, P.~G., \& Henderson, A.~P.\ 1967, \apj, 147, 471 
\bibitem[Panagia \& Felli(1975)]{P-F75} Panagia, N., \& Felli, M.\ 1975, \aap, 39, 1 
\bibitem[Rauw et al.(2007)]{Rauw07} Rauw, G., Manfroid, J., Gosset, E., Naz{\'e}, Y., Sana, H., De Becker, M., Foellmi, C., \& Moffat, A.~F.~J.\ 2007, \aap, 463, 981 
\bibitem[Russeil et al.(2007)]{Russeil07} Russeil, D., Adami, C., \& Georgelin, Y.~M.\ 2007, \aap, 470, 161
\bibitem[Samal et al.(2007)]{Samal07} Samal, M.~R., Pandey, A.~K., Ojha, D.~K., Ghosh, S.~K., Kulkarni, V.~K., \& Bhatt, B.~C.\ 2007, \apj, 671, 555 
\bibitem[Sharpless(1959)]{Sharpless59} Sharpless, S.\ 1959, \apjs, 4, 257 
\bibitem[Smith et al.(2002)]{Smith02} Smith, L.~J., Norris, R.~P.~F., \& Crowther, P.~A.\ 2002, \mnras, 337, 1309
\bibitem[Snell \& Bally(1986)]{S-B86} Snell, R.~L., \& Bally, J.\ 1986, \apj, 303, 683
\bibitem[Sternberg et al.(2003)]{Sternberg03} Sternberg, A., Hoffmann, T.~L., \& Pauldrach, A.~W.~A.\ 2003, \apj, 599, 1333 
\bibitem[Watson et al.(2008)]{Watson08} Watson, C., et al.\ 2008, \apj, 681, 1341 
\bibitem[Yun et al.(2008)]{Yun08} Yun, J.~L., Djurvik, A.~A., Delgado, A.~J. \& Alfaro, E.~J.\ 2008, \aap, 483, 209
\bibitem[Zavagno et al.(2010)]{Zavagno10} Zavagno, A., et al.\ 2010, \aap, 518, L101 
\bibitem[Zhu et al.(2009)]{Zhu09} Zhu, Q., Davies, B., Figer, D.~F., \& Trombley, C.\ 2009, \apj, 702, 929 
\bibitem[Zinchenko et al.(1998)]{Zinchenko98} Zinchenko, I., Pirogov, L., \& Toriseva, M.\ 1998, \aaps, 133, 337 

\end{thebibliography}
\end{document}